\pdfoutput=1
\documentclass[11pt,a4paper]{article}
\usepackage{jheppub,amsfonts,amssymb,enumitem,dsfont,stackengine}
\usepackage[caption=false]{subfig}
\usepackage[table]{xcolor} 
\usepackage{array} 
\usepackage{pifont}

\newcommand{\beq}{\begin{eqnarray}}
\newcommand{\eeq}{\end{eqnarray}}
\newcommand{\non}{\nonumber\\}

\DeclareMathOperator{\SU}{SU}

\newcommand{\p}{\partial}
\renewcommand{\i}{\mathrm{i}}
\renewcommand{\d}{\mathop{}\!\mathrm{d}}

\DeclareMathOperator{\tr}{tr}

\newcommand{\calE}{\mathcal{E}}

\newcommand{\wha}{\widehat{a}}

\newcommand{\cmark}{\ding{51}}%
\newcommand{\xmark}{\ding{55}}%

\title{Isospin asymmetry and neutron stars in V-QCD}

\author{Lorenzo Bartolini$^1$,}
\emailAdd{lorenzo(at)henu.edu.cn}
\author{Sven Bjarke Gudnason$^1$,}
\affiliation{$^1$Institute of Contemporary Mathematics, School of
  Mathematics and Statistics, Henan University, Kaifeng, Henan 475004,
  P.~R.~China}
\emailAdd{gudnason(at)henu.edu.cn}
\author{Matti J\"arvinen$^{2,3}$}
\affiliation{$^2$Asia Pacific Center for Theoretical Physics, Pohang
  37673, Republic of Korea} 
\affiliation{$^3$Department of Physics, Pohang University of Science
  and Technology, Pohang 37673, Republic of Korea}
\emailAdd{matti.jarvinen(at)apctp.org}

\abstract{
Isospin asymmetric nuclear matter is introduced to V-QCD, a bottom-up holographic Quantum Chromodynamics (QCD) model.  
Using a small isospin chemical potential we extract the symmetry energy in the model, finding excellent agreement with experimental results for some of the potentials. Extending the calculation for finite and arbitrary sized isospin chemical potentials, we construct $\beta$-equilibrated neutron stars via the usual Tolman-Oppenheimer-Volkov (TOV) equations. We find, pleasingly, that the  neutron stars passing the mass/radius and tidal deformability constraints are those with the potentials that also lead to excellent symmetry energies.
}

\begin{document}
\begin{flushright}
 APCTP Pre2025 - 008
\end{flushright}

\maketitle

\section{Introduction}

Quantum Chromodynamics is the established fundamental theory of the
strong nuclear interactions. Due to asymptotic freedom, a property of
non-Abelian quantum gauge theories with sufficiently little matter
content, QCD is weakly coupled at high energies, where perturbative
QCD (pQCD) becomes reliable. Unfortunately, the low-energy world of
nuclei -- protons and neutrons -- is then described by QCD at strong
coupling. 
Lattice formulation of QCD (LQCD) is a powerful numerical tool that
can calculate observables on a discretized Euclidean space (a
four-dimensional hypercube). 
Unfortunately, due to a technical problem called the sign problem \cite{deForcrand:2009zkb},
LQCD struggles to converge when chemical potentials are introduced,
which is needed for simulating QCD at finite densities. 
Neutron stars are the most compact objects in the Universe
that have not collapsed into black holes and they provide us with data on
mass, radius, and spin etc. -- but they are at the nexus of the
above-described problems: Low energies and large densities.

The AdS/CFT correspondence is a powerful tool as a mathematical
duality: It maps couplings, numbers and observables on one side of the
duality to the other side. The fascinating property of this duality in
particular, is that it also maps strong coupling to weak coupling: A
strong coupled field theory on the $d$-dimensional boundary of an anti-de Sitter
(AdS) spacetime is mapped to a weakly coupled gravitational theory in
the bulk of said $d+1$ dimensional AdS.
The main and strongest evidence for this duality was found by
Maldacena in his seminal 1997 paper \cite{Maldacena:1997re}, mapping the entire
spectrum of primary operators of the superconformal field theory,
i.e.~$\mathcal{N}=4$ 
super-Yang-Mills (sYM), to AdS in five dimensions with independence from
coordinates on a 5-sphere.
Superconformal field theories or simply conformal field theories (CFTs) have
limited usage for particle physics, with the exception of some ideas
denoted unparticle physics \cite{Georgi:2007ek}, since they describe field theories
without reference to any scale or any size. Hence, they are not
suitable for particle physics, such as QCD.

Unlike the case of $\mathcal{N}=4$ sYM theory, holographic models for
strong interactions, dubbed holographic QCD (HQCD), have not been
derived with strong evidence for the duality to hold true and it would
often be better to say that they are holographic models, than dual theories
to true QCD.
Nevertheless, they are born with a strong-weak duality built-in and are
usually constructed to assimilate the properties of strong
interactions or at least main aspects thereof. 

Generally speaking, there are two kinds of holographic QCD models:
Top-down models that are based on string theory constructions, such as
the D3/D7 model \cite{Karch:2002sh}, the Witten-Sakai-Sugimoto (WSS) model \cite{Witten:1998zw,Sakai:2004cn}, etc. They have the
advantage of having a full-fledged string theory in the UV, which
builds in some consistency in the models, although the UV theory is
usually a web of D-branes and most likely not exactly the CFT that QCD flows
to at large (infinite) energies. 
Another branch of HQCD models are called bottom-up as they are
engineered to describe QCD purely from phenomenological considerations
or even (L)QCD data.
Popular models studied in recent years fall in the classes of hard-wall
\cite{Polchinski:2001tt} or soft-wall models \cite{Karch:2006pv}.

In this paper, we will work with V-QCD \cite{Gursoy:2007cb,Gursoy:2007er,Gursoy:2008za,Gursoy:2009jd,Jarvinen:2011qe,Jarvinen:2015ofa,Arean:2012mq,Arean:2013lta,Arean:2013tja,Alho:2012mh,Alho:2013hsa,Alho:2015zua} which is a bottom-up
HQCD model that is inspired by the Veneziano limit, which is a
variation of the 't Hooft large-$N_c$ limit. More precisely, the 't
Hooft large-$N_c$ limit of gauge theories corresponds to a fixed
number of flavors with the number of colors, $N_c$ being sent to
infinity \cite{tHooft:1973alw}.
Rescaling the gauge coupling $g_{\rm YM}^2\to g_{\rm YM}^2/N_c$,
observables remain finite with $1/N_c$ corrections in the large-$N_c$
limit.
The Veneziano limit~\cite{Veneziano:1976wm} on the other hand, also sends the number of flavors to infinity as:
\beq
N_f\rightarrow \infty,\qquad N_c\rightarrow \infty,\qquad x_f\equiv\frac{N_f}{N_c} \text{ fixed},
\eeq
and imposing the same scaling for the gauge coupling as in the 't Hooft limit.
The ``V'' in the name of the model refers to the fact that the string theory inspiration of the model makes use of this limit.
The V-QCD model is, however, a bottom-up model with gluon and quark degrees of freedom, where the quarks are fully backreacted to the glue background, which corresponds formally to the Veneziano limit given above.
The V-QCD models parameters are determined by comparing observables to (L)QCD data at finite (and small) values of $N_c$ and $N_f$ in the end.  
See ref.~\cite{Jarvinen:2021jbd} for a recent review of the model.

Neutron stars have been considered previously in HQCD models, in particular in WSS with isospin asymmetry \cite{Kovensky:2021kzl,Kovensky:2021wzu,Bartolini:2023wis}, in D3/D7 without isospin asymmetry \cite{Hoyos:2016zke,BitaghsirFadafan:2019ofb,BitaghsirFadafan:2020otb}, in hard-wall without isospin asymmetry \cite{Bartolini:2022rkl}, in V-QCD without isospin asymmetry \cite{Jokela:2018ers,Chesler:2019osn,Ecker:2019xrw,Jokela:2020piw,Jarvinen:2023jbr}, and in V-QCD using a hybrid setup where the isospin dependence was borrowed from a field theory model~\cite{Demircik:2021zll,Tootle:2022pvd,Ecker:2024kzs}\footnote{A study of neutron star matter in an Einstein-Maxwell-dilaton-scalar theory includes isospin asymmetry, but fixes the proton fraction by hand \cite{Liu:2024efy}}:
Isospin symmetric matter is the case where protons and neutrons are considered on equal footing as a rough approximation. 
Neutron stars are, however, made up of mostly neutrons as the large density allows for the baryons to be unstable (i.e.~to be neutrons) because they are neutral and hence do not suffer from Coulomb repulsion.
In order to take this properly into account in HQCD, isospin asymmetry must be incorporated.
Although isospin asymmetric baryonic matter has been considered previously in HQCD \cite{Kovensky:2021ddl,Bartolini:2022gdf,Kovensky:2023mye,Bartolini:2023wis} as well as for neutron stars in the WSS model \cite{Kovensky:2021kzl,Kovensky:2021wzu,Bartolini:2023wis}, it has not been ported to V-QCD, where the fits to LQCD infuse real-world QCD phenomenology into the model, until now.

In order to consider an object as large as a neutron star with such a large density in its interior, it is often approximated as a homogeneous matter of neutrons and protons inside, where there is no tracing of where there protons or neutrons are, but simply how many there are per unit volume.\footnote{Neutron stars have also been studied in the Skyrme model, both in the homogeneous approximation \cite{Adam:2014dqa,Adam:2015lpa,Adam:2020djl,Adam:2020jts,Adam:2020yfv} as well as with baryonic crystals \cite{Adam:2023cee,Leask:2023tti}.}
This can be approximated by the so-called homogeneous Ansatz for baryonic matter in HQCD \cite{Li:2015uea} together with a chemical isospin potential, which is imposed on the boundary, according to the holographic dictionary. Apart from drastically simplifying the analysis, the use of this Ansatz circumvents phenomenological issues such as the fact that the holographic large-$N_c$ dense nuclear matter phase is a crystal~\cite{Kaplunovsky:2010eh,Jarvinen:2020xjh}, rather than a superconducting liquid.

In this paper, we port for the first time isospin asymmetric baryonic matter to the precision fit-to-LQCD V-QCD holographic model of QCD, calculate the symmetry energy as well as fully nonlinearly backreacted asymmetric baryonic matter with finite isospin for the computation of nonspinning neutron stars.
Using the TOV equations we compute the corresponding mass-radius (MR) curves for neutron stars and further compute the $\beta$-equilibrated neutral neutron/proton/electron/muon composition of the stars.

V-QCD has already been established as a precision holographic model for quite some time. A recent and necessary development in the model was the advancement of determining the Chern-Simons (CS) term \cite{Jarvinen:2022mys}. Although not important for the pure glue sector, the CS term is of utter importance for the baryon \cite{Jarvinen:2022gcc}, since the stabilization of the size and position in the holographic direction depends strongly on the CS.
This work on the CS is thus a necessary step as it determined the functional form and coefficients of some of the functions entering the CS term. We will, however, take a few of the parameters as free parameters in this work, see below.

We find, albeit a rescaling is necessary for compensating some of all the approximations done in the process of building this model, that the symmetry energy is surprisingly better than in the WSS model \cite{Bartolini:2022gdf} and the neutron stars for some of the potentials can pass all the phenomenological mass/radius/tidal deformability bounds. 
More interestingly, we find that the equations of state that give rise to good symmetry energies \emph{also} lead to neutron stars well inside the experimental constraints.

The paper is organized as follows.
In sec.~\ref{sec:model} we introduce the model and set our notation.
In sec.~\ref{sec:symm_NM} the homogeneous Ansatz and symmetric matter is reviewed.
In sec.~\ref{sec:isospin_asymmetry}, isospin asymmetric matter is introduced into V-QCD following a generalized Ansatz previously used in the WSS and the symmetry energy is presented.
In sec.~\ref{sec:quark_phase}, the possibility of a pure quark phase is considered.
In sec.~\ref{sec:neutron_stars}, we finally present our results on the neutron stars.
We conclude with a discussion in sec.~\ref{sec:conclusions}.
Some details are delegated to the appendix \ref{app:vanishing_r_c}.

\section{The model}\label{sec:model}

We consider V-QCD which is a holographic bottom-up model dual to QCD with both gluon and quark degrees of freedom and full backreaction of the quarks to the glue~\cite{Jarvinen:2011qe}.
The V-QCD model is built around two main sectors, the glue and the flavor: The glue sector is described by Improved Holographic QCD (IHQCD)~\cite{Gursoy:2007cb,Gursoy:2007er}, a bottom-up model for Yang-Mills theory inspired by five-dimensional noncritical string theory. The flavor sector is described by Tachyonic Dirac-Born-Infeld (TDBI) and Tachyonic Chern-Simons (TCS) actions arising from a pair of space-filling D4 branes~\cite{Casero:2007ae,Bigazzi:2005md}.

The holographic dictionary between the five-dimensional fields in gravity and their dual operators in QCD is the following:
\begin{itemize}
    \item The metric $g_{MN}$ is dual to the energy-momentum tensor $T_{\mu\nu}$, where Latin (Greek) indices are five (four) dimensional.
    \item The dilaton $\phi$ is dual to the $G_{\mu\nu}^a G^{\mu\nu\,a}$ operator, where $G_{\mu\nu}^a$ is the field strength of the gluons and $a$ is the adjoint color index.
    \item The tachyon $T^{ij}$, a complex matrix field with the flavor indices $i,j = 1 \ldots N_f$, is dual to the quark bilinear operator $\bar\psi^i\psi^j$.
    \item The left- and right-handed gauge fields $\big(A_M^{L/R}\big)^{ij}$ are dual to the left- and right-handed current operators $\bar\psi^{i}(1\pm\gamma_5)\gamma_\mu \psi^j$.
\end{itemize}
Note that here the first two fields are related to the gluonic degrees of freedom. They arise from the closed string degrees of freedom in string theory. The latter two fields are flavor fields, coming from the space-filling branes, and describe open string degrees of freedom.

The action for IHQCD consists of two terms,
\beq
S_{\rm IHQCD}= S_g+S_{\rm GH},
\eeq
where the first term is the five-dimensional Einstein-dilaton gravity action and the second term is the Gibbons-Hawking boundary term:
\beq \label{eq:Sgdef}
S_g=&&M_p^3 N_c^2\int\d^5x\, \sqrt{-\det{ g}}\left[R-\frac{4}{3}g^{MN}\p_M\phi\p_N\phi +V_g(\phi)\right],\\
\label{eq:SGHdef}
S_{\rm GH}=&&M_p^3 N_c^2\int\d^5x\,\sqrt{-\det{h}}\,K.
\eeq
Here $M_p$ is the five-dimensional Planck mass, $V_g$ is the dilaton potential, 
and $h$ is the pull-back of the five-dimensional metric $g$ on the four-dimensional boundary, while $R$ and $K$ are, respectively, the scalar and extrinsic curvature.
The five-dimensional metric $g$ is 
calculated using the Ansatz
\beq \label{eq:metric}
\d s^2= e^{2A(r)}\left(\frac{\d r^2}{f(r)}-f(r)\d t^2+\d\vec{x}^2\right).
\eeq
We choose the UV boundary to be at $r=0$, while for the IR behavior there are different choices, determined by of the blackening factor $f(r)$.
In this work, we will be interested in nuclear matter in the zero temperature confined phase, hence we will set $f(r):=1$.

The TDBI action is given by
\beq \label{eq:TDBI}
S_{\rm TDBI}=-\frac{1}{2} M_p^3 N_c \tr \int \d^5x\left(V_f(\phi,T^\dag T)\sqrt{-\det{\mathbf{A}}_L}+V_f(\phi,TT^\dag)\sqrt{-\det{\mathbf{A}}_R}\right), 
\eeq
with $\tr$ denoting the trace over the flavor indices and
\begin{align}
\mathbf{A}_{L,MN} &= g_{MN} + w(\phi,T^\dag T)F_{MN}^{(L)} + \frac{\kappa(\phi,T^\dag T)}{2}\left[\left(D_M T\right)^\dag\left(D_N T\right)+\left(D_N T\right)^\dag\left(D_M T\right)\right],\non
\mathbf{A}_{R,MN} &= g_{MN} + w(\phi,T^\dag T)F_{MN}^{(R)} + \frac{\kappa(\phi,T T^\dag)}{2}\left[\left(D_M T\right)\left(D_N T\right)^\dag+\left(D_N T\right)\left(D_M T\right)^\dag\right],\non
D_M T &= \p_M T +\i TA_M^L - \i A_M^R T, \non
F^{(L/R)}_{MN} & = \partial_M A^{L/R}_N- \partial_N A^{L/R}_M -\i [A^{L/R}_M,A^{L/R}_N],
\end{align}
where $\mathbf{A}_{L/R}$ are matrices in the square root of the TDBI action, not to be confused with $A_{M}^{L/R}$ which are gauge fields.

We will consider simple choices for the tachyon and dilaton dependence of the various potentials following refs.~\cite{Jarvinen:2011qe,Jokela:2018ers}: In particular, we will assume the tachyon to be proportional to the unit matrix $T=\tau(r)\mathds{1}$, as appropriate for a setup with identical quark masses (we will work with vanishing quark masses), and an exponential potential in the squared tachyon for the function $V_f$:
\beq
V_f(\phi,TT^\dag)= V_{f0}(\phi)e^{-\tau^2},
\eeq
and assume tachyon dependence to be absent for other functions: $\kappa = \kappa(\phi)$, $w=w(\phi)$.

We follow the usual approach with nuclear matter in gauge/gravity duality, and treat the gauge fields in the action as probe fields. That is, we replace eq.~\eqref{eq:TDBI} by its leading order expansion in the terms involving gauge fields~\cite{Ishii:2019gta},
\begin{align}
 S_{\rm TDBI} &\approx  S_{\rm TDBI}^{(0)}+S_{\rm TDBI}^{(1)}, \\
\label{eq:S0TDBI}
 S_{\rm TDBI}^{(0)} &= -M_p^3 N_c N_f \int\d^5x\, V_{f0}(\phi)e^{-\tau^2} \sqrt{-\det \tilde g} , \\
 S_{\rm TDBI}^{(1)} &= -M_p^3 N_c \int\d^5x\, V_{f0}(\phi)e^{-\tau^2} \sqrt{-\det \tilde g}\, \mathrm{tr}\bigg[\frac{1}{2}\kappa(\phi)\tau^2 \left(\tilde g^{-1}\right)^{MN} A_M A_N \nonumber
 \\
&\phantom{=\ }  -\frac{1}{8}w(\phi) \left(\tilde g^{-1}\right)^{MN}  \left(\tilde g^{-1}\right)^{PQ}\left(F^{(L)}_{NP}F^{(L)}_{QM}+F^{(R)}_{NP}F^{(R)}_{QM}\right)\bigg], \label{eq:S1TDBI}
\end{align}
where $A_M=A^L_M-A^R_M$ and $\tilde g$ is the effective open string metric for the background,
\beq
 \tilde g_{MN} = g_{MN} + \kappa(\phi)(\tau')^2\delta^r_M\delta^r_N .
\eeq
Expanding the action to the first nontrivial order in the gauge fields also removes an ambiguity of the DBI expression~\eqref{eq:TDBI}: As the non-Abelian gauge fields do not commute, one would need to prescribe in which order to take the trace over them. In eq.~\eqref{eq:S1TDBI} this ambiguity is resolved as there are only two flavored fields in the trace. 

Finally, for the TCS term, we employ the expression provided in ref.~\cite{Jarvinen:2022mys}, which generalizes the term derived in ref.~\cite{Casero:2007ae}. The general expression for the TCS term is complicated, including both a five-dimensional ``bulk'' term, a closed term, and a boundary term. The two latter terms are the same as those found in QCD by using the flavor anomalies~\cite{Witten:1983tw,Kaymakcalan:1983qq,Manes:1984gk}. However, we will only need the bulk term here; for the discussion of the boundary terms see ref.~\cite{Bartolini:2023eam}. Inserting the Ansatz given above, the only relevant terms turn out to be
\begin{align}
 S_{\rm TCS} &= \frac{\i N_c}{4\pi^2} \int \Omega_5^s,\\
 \Omega_5^s &= \frac{1}{24}F_1(\tau)\, \mathrm{tr}\left[A\wedge A\wedge A\wedge \left(F^{(L)}+F^{(R)}\right)\right]
 \non&\phantom{=} -\frac{\i}{24}F_3(\tau)\, \mathrm{tr}\left[A\wedge \left(F^{(L)}-F^{(R)}\right)\wedge \left(F^{(L)}-F^{(R)}\right)\right],
 \label{eq:STCS}
\end{align}
where $A=A^L-A^R$ is a 1-form and both functions $F_1$ and $F_3$ need to approach unity at the boundary for full consistency with the flavor anomalies; they should vanish at large values of $\tau$ in order to avoid unphysical IR contributions to the action~\cite{Jarvinen:2022mys}\footnote{The mapping between functions $F_i$ here and the functions $f_i$ in this reference is $F_1 = 12 (-f_1 + 2 \i f_2 - f_3)$, $F_3 = 12(f_3-f_1)$.}.

For the final choice of the functions $F_1$ and $F_3$, we take the result from the flat space analysis~\cite{Casero:2007ae}, but allow a small modification~\cite{Ishii:2019gta,Jarvinen:2022gcc} in the choice of the rescaling of the tachyon dependence, in order to better reproduce the properties of dense nuclear matter. Explicitly, this choice is given by
\beq
 F_1(\tau) = e^{-b_1\tau^2}(1-2b_1\tau^2), \qquad 
 F_3(\tau) = e^{-b_3\tau^2},
\eeq
where the positive parameters $b_1$ and $b_3$ reflect the choice of scaling.
Note that the choice of the TCS term leaves intact the 
fit of the background and the physics of the quark phase, for which the instanton number vanishes and the TCS term evaluates to zero.

To summarize, the full action governing the background (metric, dilaton and the tachyon) is then given by 
\beq 
S_\mathrm{bg}\equiv S_g+S_{\rm GH}+S_{\rm TDBI}^{(0)}
\eeq
in eqs.~\eqref{eq:Sgdef},~\eqref{eq:SGHdef}, and~\eqref{eq:S0TDBI}, whereas the probe gauge fields are described by 
\beq \label{eq:Sgf}
 S_\mathrm{gf}\equiv S_{\rm TDBI}^{(1)}+S_{\rm TCS}
\eeq
in eqs.~\eqref{eq:S1TDBI} and~\eqref{eq:STCS}.

To build hybrid equations of state (EOS) we need to choose a procedure to match the holographic and low-energy equations. We will present two procedures different in spirit, which we will refer to as procedures (a) and (b) from now on: 
\begin{enumerate}[label=(\alph*)]
    \item Choose a phenomenologically reasonable density $n_t$ at which to perform the matching, and introduce a rescaling parameter $c_b$ of the flavor action to obtain realistic pressure scales by replacing $S_{\rm TDBI}^{(1)} \mapsto c_b S_{\rm TDBI}^{(1)}$ and $S_{\rm TCS} \mapsto c_b S_{\rm TCS}$. We then choose $b_1,c_b$ in order to obtain an equation of state with continuous pressure, baryon number density and energy density. This is exactly the same approach followed in refs.~\cite{Ecker:2019xrw,Jokela:2020piw}, and essentially introduces a new parameter in an attempt to correct for the various approximations of the holographic model. This procedure tends to produce the most phenomenologically reasonable equations of state, at the price of losing predictability and control over the areas of the model that would require refinement.
    \item Fit the model to properties of nuclear matter at saturation, then determine the transition between the two equations of state by requiring continuity of pressure and baryon number density. This is exactly the same approach as that followed in ref.~\cite{Bartolini:2023wis}. To do so, we fit 
    $b_1$ to have the model-determined baryonic saturation density $n_0$ coincide with the phenomenological value $n_S=0.16\rm fm^{-3}$, 
    This procedure tends to produce less reliable equations of state, with some unphysical features, such as a first-order phase transition at densities around saturation, while focusing on retaining control over the qualitative features of the holographic method, in particular highlighting the degree of failure of the homogeneous Ansatz at lower densities.
\end{enumerate}

For both procedures, we will use the holographic equation of state obtained with two different background choices, commonly labeled 5b and 7a (see App.~A of ref.~\cite{Jokela:2020piw} and App.~B of~ref.~\cite{Ishii:2019gta} for details on the definitions). These models were determined primarily by comparing to available the lattice data for the thermodynamics of QCD. There are also other fits of the V-QCD model available~\cite{Amorim:2021gat,Jarvinen:2022gcc}, which also take into account data for particle spectra, but produce a less precise description of the finite temperature lattice data. For low-energy equations of state, we use the three variants from Hebeler-Lattimer-Pethick-Schwenk \cite{Hebeler:2013nza}, and SLy4 \cite{Douchin:2001sv,Haensel:1993zw}. We will perform procedure (a) for three values of $n_t$, namely $\{1.2,1.5,1.8\} n_S$, for a total of 30 different equations of state, where $n_S$ is the phenomenological saturation density (the soft and intermediate EOS from ref.~\cite{Hebeler:2013nza} result in a single hybrid EOS for procedure (b)).

\section{Symmetric nuclear matter}\label{sec:symm_NM}

To describe dense nuclear matter we employ the framework usually referred to as the ``homogeneous Ansatz''. It consists in assuming the gauge fields only depend on the holographic coordinate $r$: This can be thought of as being the result of smearing a large  number of instantons (each carrying unit baryon number) over $\mathbb{R}^3$, and the field configuration inherits many similarities with the single instanton in singular gauge, despite a rigorous derivation is still not available.

The first step is to set $N_f=2$ and employ the only consistent homogeneous Ansatz for the fields $A^i_{L/R}$, given by~\cite{Rozali:2007rx,Li:2015uea,Bartolini:2022rkl,Ishii:2019gta}:
\beq\label{eq:ansatzH}
A_i^L=-A_i^R= -\frac{H(r)}{2}\sigma^i,
\eeq
with $\sigma^i$, $i=1,2,3$ being the three Pauli matrices.
The other field that is turned on in the isospin-symmetric configuration is the Abelian potential $A_t(r)$. 
The only allowed structure for it is then
\beq\label{eq:ansatzhata0}
A_t^L=A_t^R= \frac{\wha_0(r)}{2}\mathds{1}.
\eeq
With this normalization, the UV value of this field is holographically dual to the quark chemical potential, and it relates to the 
baryon chemical potential $\mu_B$ that couples to the baryon number density as
\beq
\frac{1}{2}\wha_0(r=0)\equiv \mu = \frac{1}{N_c} \mu_B .
\eeq

To compute the baryon number for this configuration (together with the tachyon), we integrate the equation of motion for $\widehat a_0$ obtained from the gauge field action in eq.~\eqref{eq:Sgf}. We find~\cite{Ishii:2019gta} 
\begin{align}
N_B&= -\frac{1}{4\pi^2}\int\d^3x\d r \frac{\d}{\d r}\left[ F_1(\tau(r)) H(r)^3    \right].
\end{align}
We see that $N_B$ vanishes for every smooth profile $H(r)$ since $H(r=0)=0$ to have a finite energy configuration \cite{Rozali:2007rx}.
A way to retain both a finite nonvanishing baryon number and a homogeneous field configuration is to allow
the function $H(r)$ to have discontinuities \cite{Li:2015uea}: The simplest choice is for it to have a discontinuity in the bulk at some finite $r=r_c$ (that will be determined by extremizing the action), and have $H(r>r_c)=0$. This roughly corresponds to placing the instantons at $r=r_c$ before the smearing procedure. 
Analogously to the baryons/instantons forming a lattice in the holographic direction following ``popcorn transitions''~\cite{Kaplunovsky:2012gb,Kaplunovsky:2013iza}, the function $H(r)$ can in principle have multiple discontinuities (see ref.~\cite{CruzRojas:2023ugm}), a refinement we do not include in the present work.

We do, however, include the dynamical determination of the location of $r_c$, which moves toward the UV as the baryonic density increases. A possible physical interpretation of this quantity has been given in ref.~\cite{Rozali:2007rx} as follows: The holographic coordinate $r$ is dual to the energy scale of the dual QFT, so that the homogeneous nuclear-matter distribution in this direction would be dual to the spectrum of energies of the condensed baryons. Given that the distribution has a sharp edge at $r_c$, this can be interpreted as being dual to the location of a Fermi surface. In ref.~\cite{Gorsky:2015pra} it was observed in the context of a hard-wall model that the increase in the value of the chiral condensate leads to the movement of the baryon's location from the IR towards the UV. Given that increasing the chiral condensate increases the energy (i.e.~mass, in the case of a single instanton) of the baryon, this result is also consistent with the interpretation that $r_c$ should be dual to the typical energy scale of the baryon. Note that, in the model we are considering, the function $F_1(\tau)$ has a node at some $r_n$ that solves $\tau (r_n)=\frac{1}{\sqrt{2 b_1}}$. This implies that we can never have $r_c<r_n$, and so $r_c$ cannot shrink arbitrarily far from $r_c\sim\frac{1}{\Lambda_\mathrm{QCD}}$. 

Other components of the gauge field vanish due to the absence of sources, and we are then left with two ordinary differential equations (ODEs) for $H(r),\widehat{a}_0(r)$ to
be solved numerically, which we choose to do using a shooting method. A convenient choice is to shoot from the IR, since we can then 
fix the baryon number density $n_B$ and use it to impose the Dirichlet boundary condition
\beq
H(r=r_c)^3 = -\frac{ 4\pi^2  n_B}{F_1(\tau(r_c))}.
\eeq
The vanishing of boundary terms in the variation of the action imposes Neumann boundary conditions for $\wha_0(r_c)$, and fixes the derivative $H'(r_c)$ as a function of $H(r_c)^2$ \cite{Bartolini:2023eam}:
\beq
4 \pi^2  M_p\left(\frac{e^{2 A(r)} V_f(\phi,\tau) w(\phi)^2}{\sqrt{e^{2 A(r)} + \kappa(\phi)\tau'^2}} H'\right)_{r=r_c}= \left(\wha_0 F_1(\tau)H^2\right)_{r=r_c}.
\eeq
The value of $\wha_0(r_c)$ necessary to impose $H(r=0)=0$ thus becomes the only shooting parameter. The procedure of determining $r_c$ is made easy by this choice of shooting method: The variation of the on-shell action 
with respect to $r_c$ in our setup with a single discontinuity and vanishing fields for $r>r_c$, is given by
the Lagrangian density evaluated at $r_c$. It thus suffices to  numerically solve the algebraic equation $\mathcal{L}(r=r_c)=0$ for each choice of shooting parameter.

The saturation density $n_0$ can be derived from the model by finding the onset of baryonic matter at which the grand potential for nuclear matter vanishes, $\Omega(n_0)=0$. Throughout the rest of this work, we will keep separated the concepts of the model-derived saturation density $n_0$ and the  phenomenological saturation density $n_S$. The value of the former is determined by the choice of $b_1$ and is subject to an intrinsic error introduced by the assumption of homogeneity, while the value of the latter is the observed one of $n_S=0.16\rm fm^{-3}$.
When following procedure (b), the value of $b_1$ is chosen by requiring $n_0=n_S$, while for procedure (a) the two values in general do not coincide. Since procedure (a) is developed with the scope of matching the holographic EOS with other phenomenological ones at densities higher than saturation, then it is not a severe problem that in this case $n_0\neq n_S$. 

\section{Isospin asymmetry}\label{sec:isospin_asymmetry}

The introduction of isospin asymmetry is performed by turning on a nonvanishing isospin chemical potential $\mu_I$. This is done according to the standard holographic dictionary by imposing the UV boundary condition on $\big(A_t^{L/R}\big)^a$:
\beq
A_t^L(r=0)=A_t^R(r=0)=\frac{\mu_I}{2}\sigma^3,
\eeq
where we have aligned the group direction $\sigma^3$ with the third component of the isospin.

The boundary condition turns on components of the gauge field that could be set to zero in the symmetric case, in particular it turns on $\big(A_t^{L/R}\big)^{a=3},\widehat{A}_3^{L/R}$ and partially removes the symmetry of the field $A_i^{L/R}$. The new self-consistent Ansatz for the isospin asymmetric setup is then given by:
\beq
&&A_i^L=-A_i^R= -\frac{H(r)}{2}\sigma^i ,\qquad\qquad \text{with } i=1,2 \label{eq:IsospinAi}\\
&&A_3^L=-A_3^R= -\frac{H_3(r)}{2}\sigma^3 - \frac{L_3(r)}{2} \mathds{1},\label{eq:IsospinA3}\\
&&A_t^L=A_t^R= a_0(r)\sigma^3 + \frac{\widehat{a}_0}{2} \mathds{1}.\label{eq:IsospinA0}
\eeq
The other components allowed in principle by the symmetries (the Abelian component for $A_1,A_2$ and the missing non-Abelian components $A_t^1,A_t^2$, on top of terms involving the antisymmetric symbol $\epsilon^{ij}$) are unsourced and vanishing due to the equations of motion.
The equations of motion for the fields $A_r^L,A_r^R$ are satisfied by their trivial solution, $A_r^L=A_r^R=0$, while the tachyon $\tau(r)$ is kept fixed to its background configuration (i.e.~it satisfies its equation of motion in the absence of baryons), as the flavor gauge fields are taken to be in the probe approximation.

The structure of this Ansatz can be thought of as originating from the inhomogeneous configuration as follows: The holographic baryon is an instanton-like soliton, whose topological number arises from its mapping $S^3\rightarrow S^3$. The field configuration for the baryon/instanton in V-QCD is constructed in ref.~\cite{Jarvinen:2022gcc}, and it is the general Ansatz that can be written by imposing cylindrical symmetry together with parity and time-reversal invariance. The process of smearing the baryon on $\mathbb{R}^3$ then eliminates every term that is odd in $x^i$ (the coordinate on $\mathbb{R}^3$). The topological number that was originally introduced by the wrapping of $S^3$ on $S^3$ can now only be retained by the presence of a discontinuity in the fields. This picture holds both for a static instanton, in which case we obtain the Ansatz in eqs.~\eqref{eq:ansatzH}-\eqref{eq:ansatzhata0}, and for a rotating soliton, in which case we recover the Ansatz in eqs.~\eqref{eq:IsospinAi}-\eqref{eq:IsospinA0}, as appropriate. We obtain the same exact Ansatz if we start from the most general one compatible with the symmetries of the homogeneous configuration, and then carefully eliminate all unsourced fields.

The UV boundary conditions are again obtained by requiring the vanishing of the sources on the field theory side except for the chemical potentials, so we have
\beq\label{eq:UVbdygeneral}
&&H(r=0)=H_3(r=0)=L_3(r=0)=0,\\
&&\widehat{a}_0(r=0)=2\mu,\\
&&a_0(r=0)=\frac{\mu_I}{2}.
\eeq
The IR boundary conditions instead are read off the boundary terms that remain when performing the variation of the action and imposing the equations of motion:
\begin{align}
\widehat{a}'_0(r_c)=0,\qquad
L'_3(r_c)&=0,\label{eq:bdya0hatANDL3}\\
24 \pi^2 M_p^3 \left(\frac{e^{2 A(r)} V_f(\phi,\tau) w(\phi)^2}{\sqrt{e^{2 A(r)} + \kappa(\phi)\tau'^2}} a'_0\right)_{(r=r_c)}&=\left[\left(F_1(\tau)+2F_3(\tau)\right)H^2L_3\right]_{r=r_c},\label{eq:bdya0}\\
4 \pi^2  M_p^3\left(\frac{e^{2 A(r)} V_f(\phi,\tau) w(\phi)^2}{\sqrt{e^{2 A(r)} + \kappa(\phi)\tau'^2}} H'\right)_{r=r_c}&= H\big(\wha_0 F_1(\tau)H_3-2a_0 F_3(\tau)L_3\big)_{r=r_c},\label{eq:bdyH}\\
4 \pi^2  M_p^3\left(\frac{e^{2 A(r)} V_f(\phi,\tau) w(\phi)^2}{\sqrt{e^{2 A(r)} + \kappa(\phi)\tau'^2}} H_3'\right)_{r=r_c}&= \left(\wha_0 F_1(\tau)H^2\right)_{r=r_c},\label{eq:bdyH3}
\end{align}
with the two equations in \eqref{eq:bdya0hatANDL3} coming from the variation of the action with respect to $\widehat{a}_0(r)$ and $L_3(r)$, respectively, and equations \eqref{eq:bdya0}, \eqref{eq:bdyH} and \eqref{eq:bdyH3} are due to the variation of the action with respect to $a_0(r)$, $H(r)$ and $H_3(r)$, respectively.

To work at a fixed baryonic density we impose the condition 
\beq
H_3(r_c) H(r_c)^2 = -\frac{ 4\pi^2  n_B}{F_1(\tau(r_c))},
\eeq
which in the symmetric case was enough to fix the IR value $H(r_c)$, but now only provides a relation between $H_3(r_c)$ and $H(r_c)$. For the shooting method we use the parametrization $H_3(r_c)=\delta_H H(r_c)$, where we have introduced the parameter $\delta_H$ being the ratio of the third and the other components on the boundary. 
In the symmetric scenario $\delta_H=1$, but in the asymmetric case it is treated as a shooting parameter that is fixed by the conditions~\eqref{eq:UVbdygeneral}.

\subsection{Small isospin approximation: Symmetry energy}

\begin{figure}[!ht]
  \centering
 \includegraphics[width=0.75\textwidth]{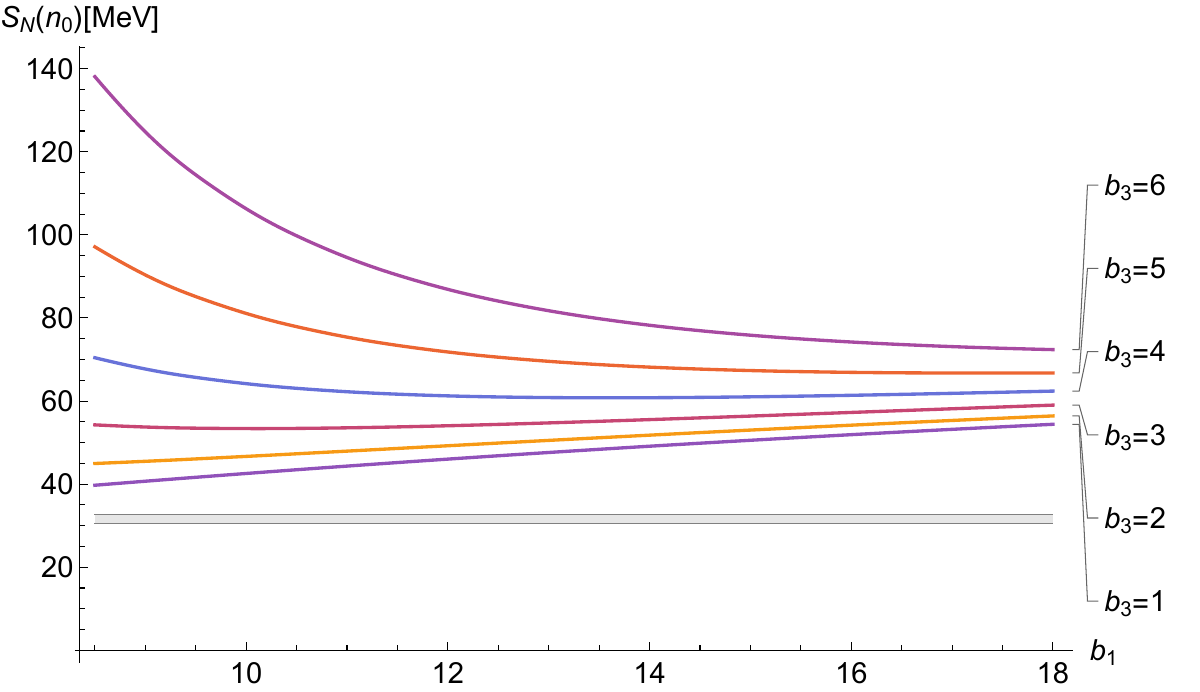}
  \caption{Symmetry energy at saturation density $n_0$ (as determined by the model) as a function of $b_1$ for various values of $b_3$. In gray the phenomenologically measured value including its experimental error.}
  \label{fig:esymmetry_b1}
  \end{figure}
  
We will now compute the symmetry energy of the system. For homogeneous nuclear matter it is defined as the coefficient of the quadratic term in the expansion of the energy per nucleon in the asymmetry parameter $\beta=(N-Z)/A$:
\beq\label{eq:SNdef}
\frac{\mathcal{E}}{n_B}(n_B,\beta)= \frac{\mathcal{E}_0}{n_B}(n_B,0) + S_N(n_B) \beta^2 +\cdots
\eeq
It is indeed possible to compute $S_N(n_B)$ from the above expansion by using that $n_I=\frac{\p \mathcal{E}}{\p \mu_I}$: 
\beq\label{eq:SNderivative}
S_N(n_B)=\frac{n_B}{8}\left.\frac{\p \mu_I}{\p n_I}\right|_{n_I=0},
\eeq
as well as the nuclear physics convention assigning isospin number $+\frac{1}{2}$ $(-\frac{1}{2})$ to the proton (neutron).

\begin{figure}[!ht]
  \centering
 \stackon[5pt]{\includegraphics[width=0.49\textwidth]{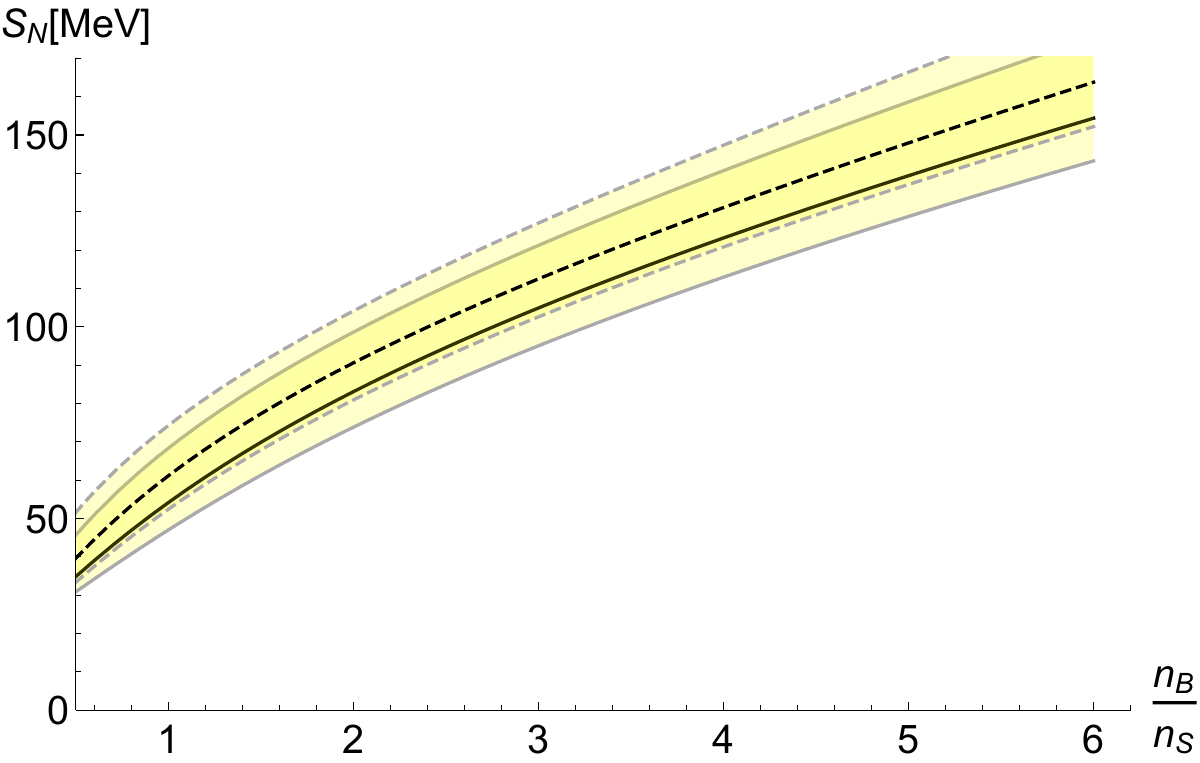}}{\footnotesize Procedure (b)}
 \stackon[5pt]{\includegraphics[width=0.49\textwidth]{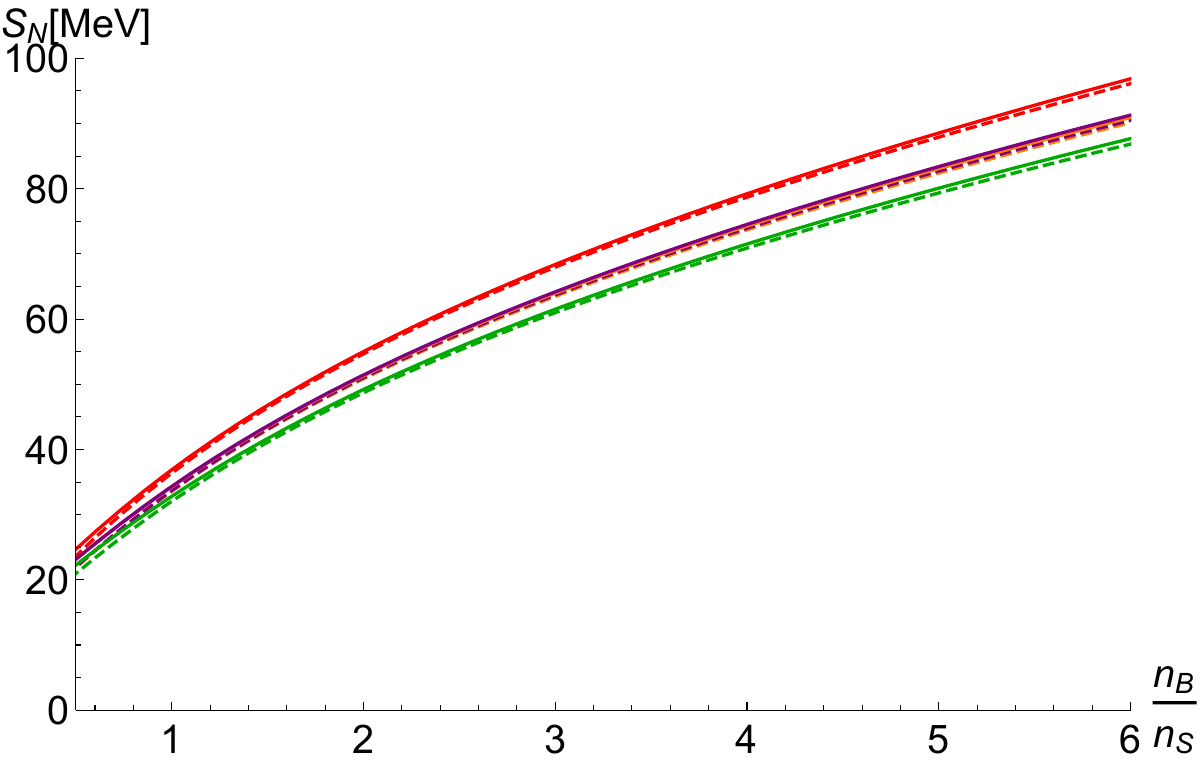}}{\footnotesize Procedure (a) $n_t=1.5n_S$}\\[10pt]
 \stackon[5pt]{\includegraphics[width=0.49\textwidth]{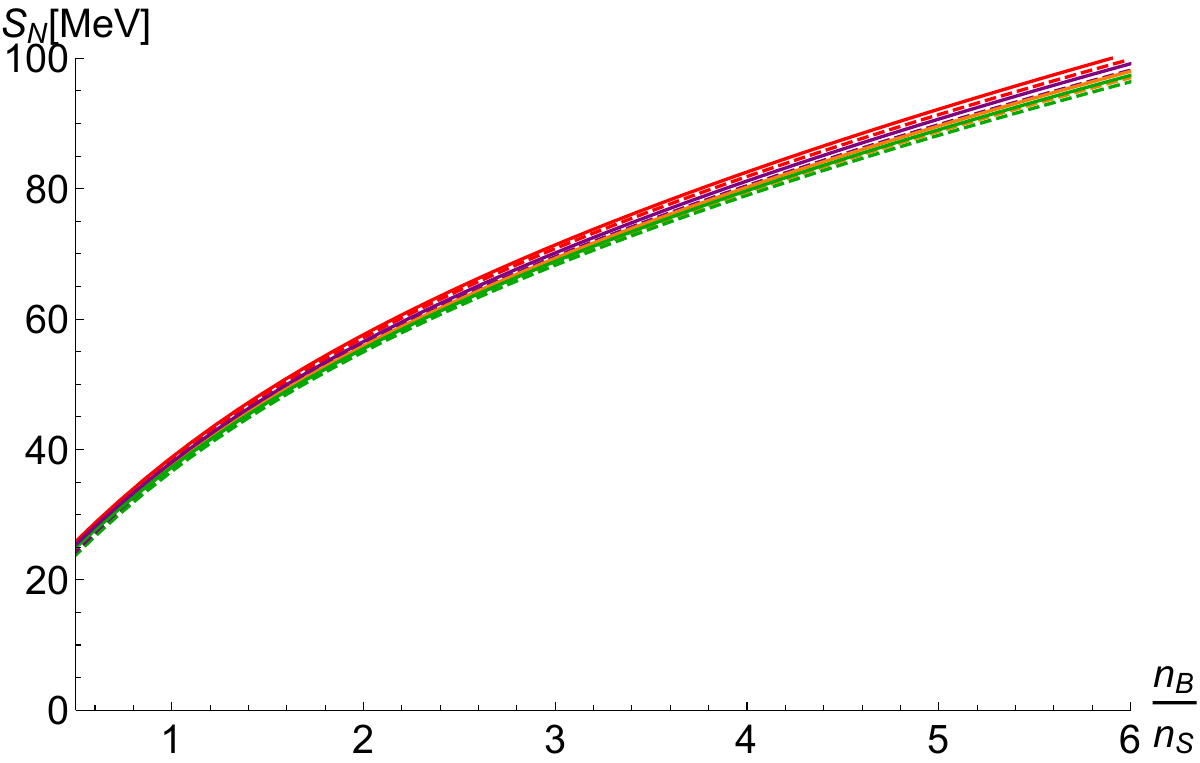}}{\footnotesize Procedure (a) $n_t=1.2n_S$}
\stackon[5pt]{ \includegraphics[width=0.49\textwidth]{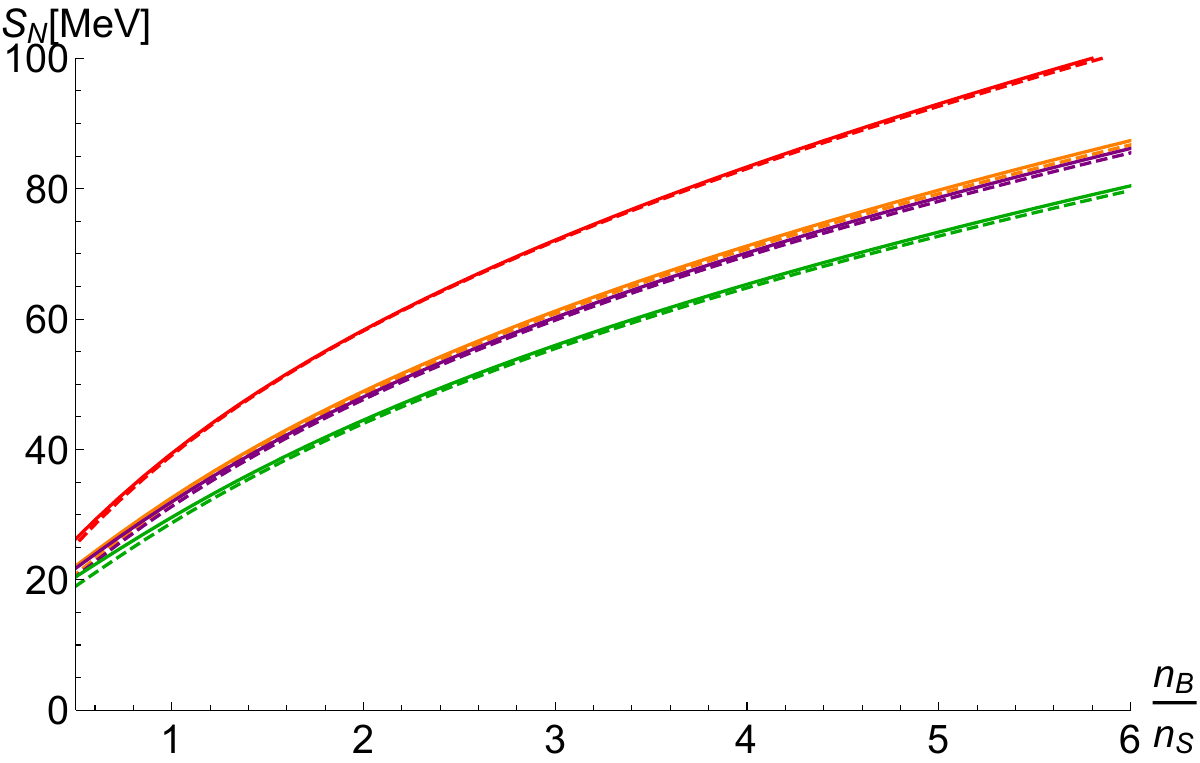}}{\footnotesize Procedure (a) $n_t=1.8n_S$}
  \caption{The symmetry energy as a function of density measured in units of the saturation density $n_S$. Top left panel: Matching procedure (b): A yellow band spanning the values $b_1\in \left[12,22\right]$ with gray boundaries is presented; the black curves correspond to the physical saturation density.
  Remaining panels: Matching procedure (a) with matching density $n_t$. 
  The color coding is: Green for HLPS soft, orange for HLPS intermediate, red for HLPS stiff, purple for SLy4. Solid (dashed) lines correspond to the choice of V-QCD potentials 7a (5b).}
  \label{fig:esymmetry}
  \end{figure}

It is, however, more convenient to exploit the fact that the computation has to be done around the isospin symmetric configuration, so that we can linearize the $\mu_I$-dependence of the fields and drop higher-order corrections. Let us explicitly manifest the linear $\mu_I$ dependence of the fields via the expansion:
\begin{align}
a_0(r) &= \widetilde{a}_0(r) \mu_I + \mathcal{O}(\mu_I^2),\\
L_3(r) &=\widetilde{L}_3(r) \mu_I+ \mathcal{O}(\mu_I^2),\\
H_3(r) &= H(r) + \mathcal{O}(\mu_I^2),
\end{align}
where the functions $\widehat{a}_0(r)$ and $H(r)$ receive corrections only at order $\mathcal{O}(\mu_I^2)$ (consistent with the equations of motion).\footnote{This setup is equivalent to that of ref.~\cite{Bartolini:2022gdf}. The approximation of small $\mu_I$ is equivalent to that of slow and rigid rotation in $\SU(2)$. The two descriptions are in fact connected by a gauge transformation; see App.~A of ref.~\cite{Bartolini:2022gdf}. }

Plugging this expansion into the action, we have:
\beq\label{eq:OmegasmallmuI}
\Omega= -\mathcal{E}_0(n_B) - \frac{1}{2}\Lambda (n_B) \mu_I^2,\qquad n_I=-\frac{\p \Omega}{\p \mu_I} = \Lambda(n_B)\mu_I,
\eeq
where we have introduced the functions $\mathcal{E}_0(n_B)$ and $\Lambda(n_B)$ defined by the integrals
\begin{align}
\mathcal{E}_0(n_B)\equiv M_p^3 N_c&\int^{r_c}_0 \d r\; \frac{ V_f(\phi,\tau)}{4\sqrt{e^{2 A(r)} + \kappa(\phi)\tau'^2}}\Big[ 3\left(e^{2A(r)}+\kappa(\phi)\tau'^2 \right)w(\phi)^2 H^4 \non
 &+12e^{2A(r)}\kappa(\phi)\tau^2\left(e^{2A(r)}+\tau'^2  \right) H^2 + e^{2A(r)}w(\phi)^2\left(3 H'^2 + \widehat{a}_0'^2\right)
 \Big],
\end{align}
\begin{align}
\Lambda(n_B)\equiv  2 M_p^3 N_c \int^{r_c}_0 \d r\;  & \frac{ V_f(\phi,\tau)}{4\sqrt{e^{2 A(r)} + \kappa(\phi)\tau'^2}}\left[
e^{2 A(r)} w(\phi)^2  \left(4 \widetilde{a}_0'^2+\widetilde{L}_3'^2\right)\right. \non
&\left.\mathop+4\left(e^{2 A(r)} + \kappa(\phi)\tau'^2\right)\left(2 w(\phi)^2H^2\widetilde{a}_0^2+e^{2 A(r)} \kappa(\phi)\tau^2\widetilde{L}_3^2\right)\right].\label{eq:Lambdadef}
\end{align}
With the definition of $n_I$ obtained in eq.~\eqref{eq:OmegasmallmuI} we can readily obtain the expression for the symmetry energy via eq.~\eqref{eq:SNderivative}:
\beq
S_N(n_B)=\frac{n_B}{8 \Lambda (n_B)}.
\eeq
Moreover, since $n_I=\Lambda \mu_I$ due to the above arguments, and since $n_I$ also has to be determined by the near-boundary behavior of the field $\widetilde{a}_0$ due to the field-operator holographic map, we conclude that it is not necessary to explicitly compute the integral \eqref{eq:Lambdadef}, but it is sufficient to compute $\Lambda(n_B)$ as
\beq
\Lambda(n_B) = -M_p^3 N_c\left.\left(\frac{e^{2 A(r)} V_f(\phi,\tau) w(\phi)^2}{\sqrt{e^{2 A(r)} + \kappa(\phi)\tau'^2}}   \widetilde{a}_0'\right)\right|_{r\rightarrow0}.
\eeq
In this analysis, we have not considered the possibility of a contribution to $S_N$ arising from the change of the location at which the discontinuity is placed, induced by $\mu_I$; that is, we determine the position in the bulk by solving $L(r_c)=0$ at the zeroth order in $\mu_I$. 
It can be proved that the perturbation of the Lagrangian induces a perturbation of $r_c$ quadratic in $\mu_I$, so one may wonder if this perturbation allows the unperturbed Lagrangian to contribute to $S_N$: 
It can, however, be shown by using the definition of $r_c$ that this contribution vanishes due to the equations of motion. Thus, the isospin-induced correction to $r_c$ can only contribute to a higher order in $\mu_I$, see App.~\ref{app:vanishing_r_c}.

Near saturation density it is useful to define the parameters $S_0$, $L$, $K_{\rm sym}$:
\beq
S_N(n_B) = S_0 + \frac{1}{3}L\,\frac{n_B-n_S}{n_S} + \frac{1}{18}K_{\rm sym}\left(\frac{n_B-n_S}{n_S}\right)^2 +\cdots
\eeq

We have explored the symmetry energy's ($S_N(n_0)$) dependence on $b_1$ and $b_3$ at $c_b=1$, summarized in fig.~\ref{fig:esymmetry_b1}. 
$S_N(n_0)$ is an increasing function of $b_3$ at fixed $b_1$. Moreover, in the range of $b_1$ relevant to our matching procedures, $S_N(n_0)$ is already overestimated for the flat space expression of the TCS term~\cite{Casero:2007ae} corresponding to $b_3=1$, that is, the physics of isospin asymmetry is better described for small values of $b_3$. This is in stark contrast with the preferred values of $b_1$, which governs instead the onset of nuclear matter. 
A minimal choice in modifying the flat space expression ($b_1=b_3=1$) for $S_{\rm TCS}$ would be to choose a single rescaling for both functions $F_1$, $F_3$, so that $b_1=b_3=b$. However, this would lead to a very large symmetry energy for both of our matching procedures (and consequently very high proton fractions for the nuclear matter in $\beta$-equilibrium). 
We have thus for phenomenological reasons fixed $b_3=1$, so that the rescaling only affects the function $F_1(\tau)$,  to keep the number of free parameters in the model to a minimum.
Without the introduction of the rescaling parameter $c_b$ (as in procedure (b)), this still leads to large symmetry energies, 
but remarkably the results for procedure (a) turns out to be close to (and in fact often compatible with) phenomenology\footnote{ It is possible that an appropriate modification of the functional form of $F_3$ (and $F_1$) may make the model reproduce phenomenological values of the symmetry energy as well as baryon onset, \emph{without} the introduction of the rescaling parameter $c_b$. This will be beyond the scope of the present paper. }.

In fig.~\ref{fig:esymmetry} is shown the symmetry energy, $S_N(n_B)$, for each combination of matching procedure, potential, and parameters $c_b$ and $b_1$, which is computed as the coefficient of the expansion around saturation density (see Tab.~\ref{tab:parameters} for the values of the parameters for the matching procedure (a)).
The symmetry energy at saturation density resulting from matching procedure (a) lies in the range $S_0\in[28.7,39.4]$ MeV. Breaking down this result in the various EOSs, we find that the soft one has $S_0\in[28.7,37.3]$ MeV, the intermediate one has $S_0\in[31.9,37.6]$ MeV, the one matched with SLy4 has $S_0\in[31.3,38.0]$ MeV and the stiff one has $S_0\in[36.4,39.4]$ MeV. 
The lower (upper) bounds always correspond to EOS whose holographic part is built with the 5b (7a) background.
Moreover, the parameter $L$ of the expansion around saturation lies in the range $L\in[51.6,71.0]$ MeV, with the upper (lower) bound corresponding again to the EOS from the 5b (7a) background. Both values are in good agreement (on the larger side) with phenomenology, a nontrivial result given the matching procedure: The rescaling with $c_b$ not only provides continuous EOSs, but also rescales the symmetry energy and its derivative with respect to the density to their phenomenological values. 
For the quantity $K_{\rm sym}$, the model predicts values $K_{\rm sym}\in[-106.1,-42.9]$ MeV.

A very interesting result that we will obtain in the following sections is that the EOSs that have $S_0$ close to phenomenology are also those that result in neutron stars more prone to satisfying the observational bounds. This is also a nontrivial result and may hint that the rescaling procedure is not as bad a correction to the many approximations employed here as it may seem.

\subsection{General configuration: \texorpdfstring{$\beta$}{beta}-equilibrated neutral matter}

The approximation presented in the previous section is only reliable in the limit of small isospin chemical potential: This is enough to compute the symmetry energy, but not to determine the equation of state for $\beta$-equilibrated matter. The regimes of $\beta$-equilibrium and charge neutrality tend, in fact, to produce neutron-rich matter, which then translates into a regime of large (negative in our conventions) isospin.

If our goal is to describe neutron-rich matter in the core of neutron stars, we need to solve the full isospin-dependent problem described by the Ansatz in eqs.~\eqref{eq:IsospinAi}, \eqref{eq:IsospinA3} and \eqref{eq:IsospinA0}.

While in the general configuration baryon and isospin number densities are independent, we impose $\beta$-equilibrium in the system by introducing negatively charged leptons. This introduces a new parameter to the problem, the chemical potential of leptons $\mu_l$ with $l=e,\mu$ for electrons and muons. The system is then constrained again to a single degree of freedom by imposing electric charge neutrality.

\begin{figure}[!ht]
  \centering
  \stackon[5pt]{\includegraphics[width=0.49\textwidth]{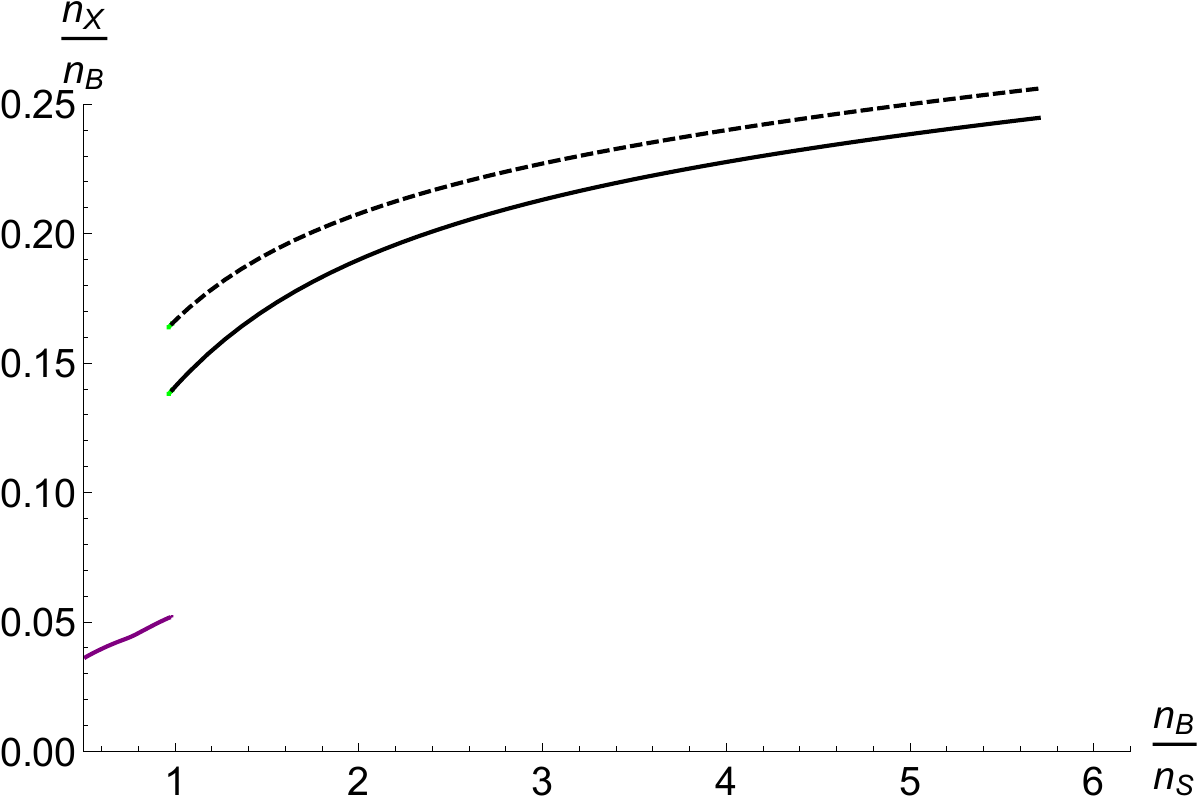}}{\footnotesize Procedure (b)}
  \stackon[5pt]{\includegraphics[width=0.49\textwidth]{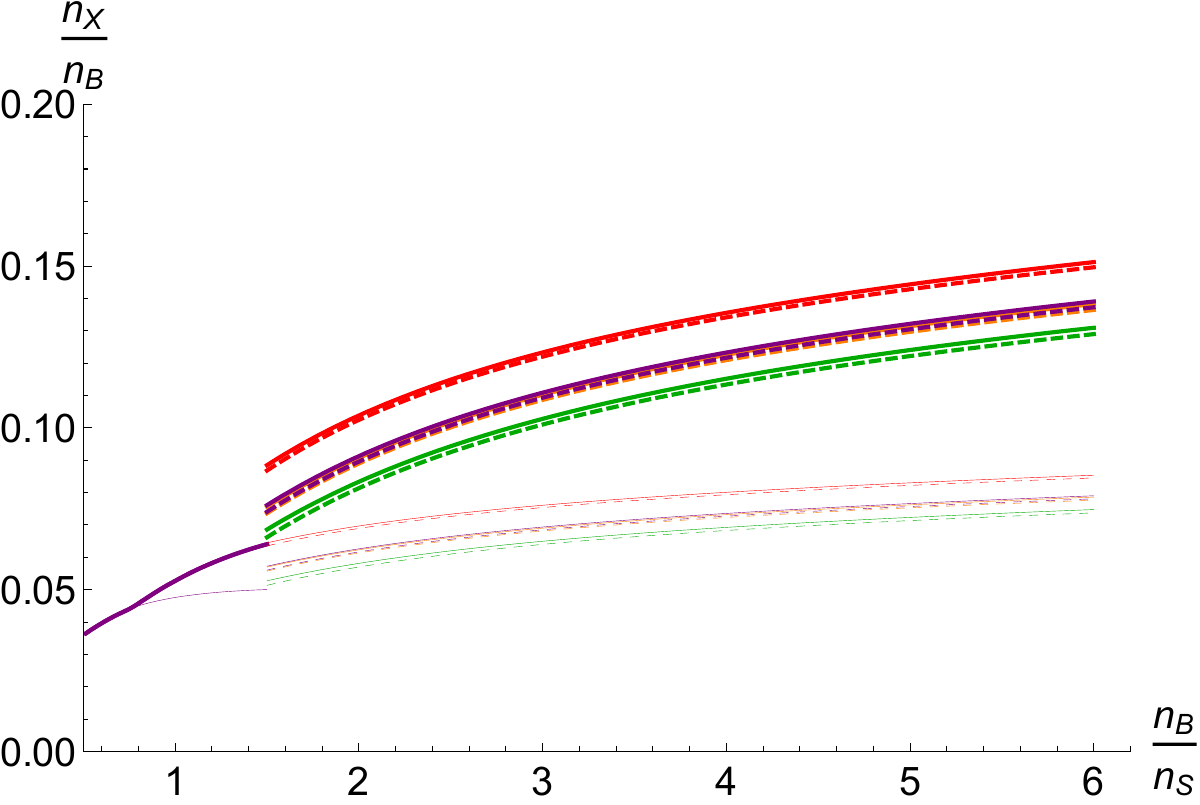}}{\footnotesize Procedure (a) $n_t=1.5n_S$}\\[10pt]
  \stackon[5pt]{\includegraphics[width=0.49\textwidth]{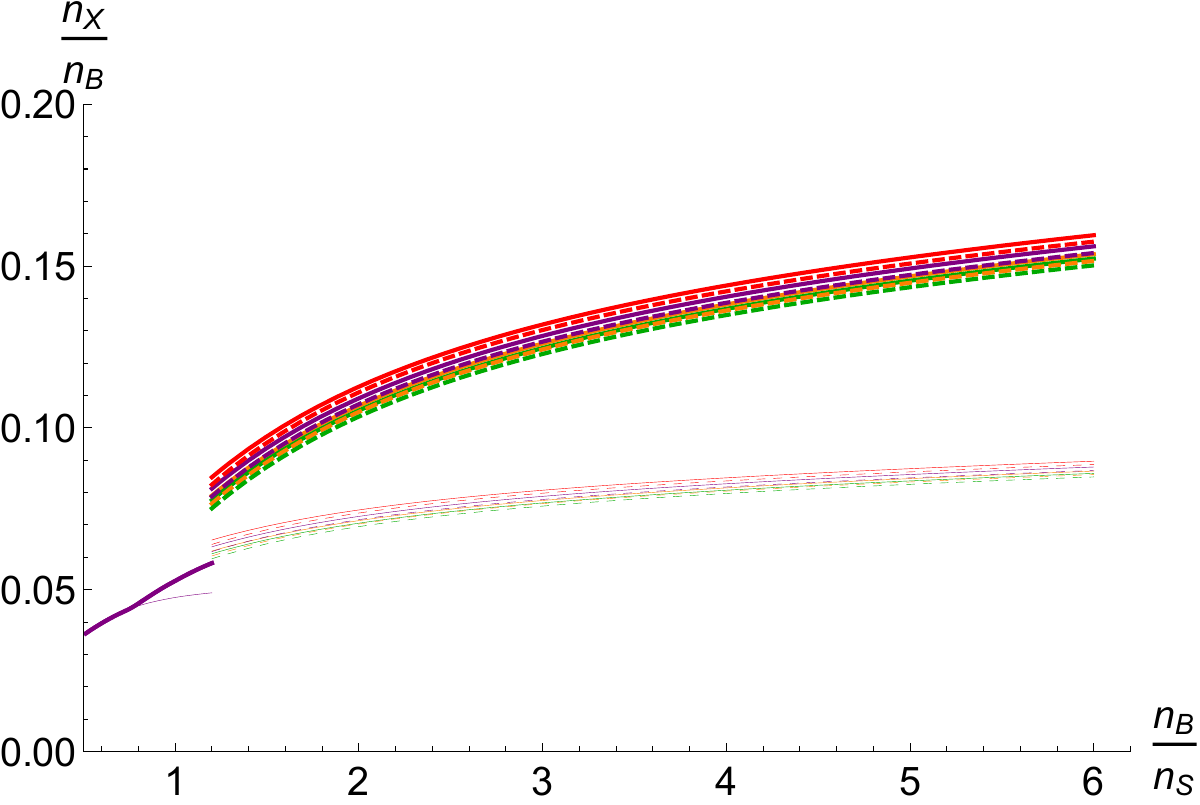}}{\footnotesize Procedure (a) $n_t=1.2n_S$}
  \stackon[5pt]{\includegraphics[width=0.49\textwidth]{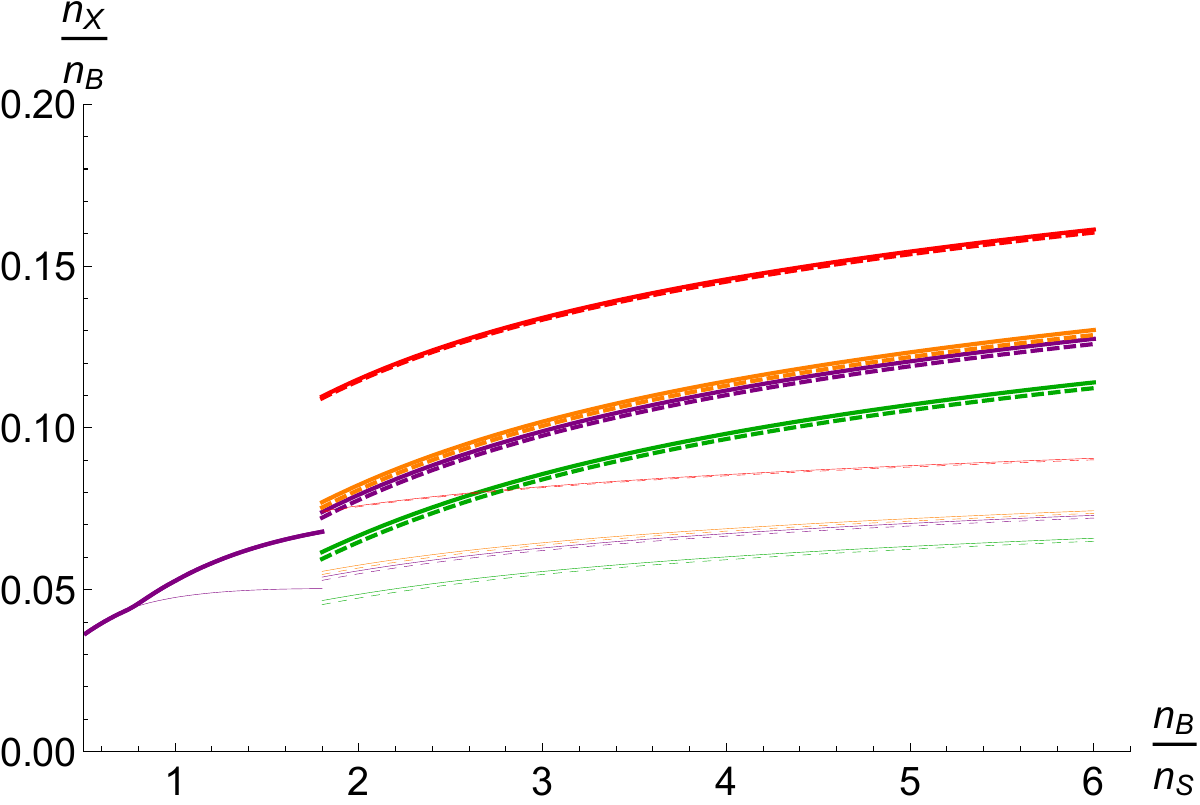}}{\footnotesize Procedure (a) $n_t=1.8n_S$}
  \caption{Proton (thick lines) and electron fractions (thin lines) as functions of baryon number density, $n_B$, in units of the saturation density $n_S$. The muon fraction is inferred from charge neutrality as the difference between the electron and proton fractions, and the neutron fraction is obtained by subtracting the proton fraction from unity. Color coding denotes the low-density equations of state used for matching: Green for HLPS soft, orange for HLPS intermediate, red for HLPS stiff, and purple for SLy4. Solid (dashed) lines correspond to the V-QCD potential 7a (5b). The top-left panel displays results from matching procedure (b), whereas the remaining panels correspond to procedure (a) for transition densities $n_t = \{1.2,1.5,1.8\}n_S$. Purple curves show the SLy4 predictions at densities below the matching point for reference.}
  \label{fig:proton_fraction}
  \end{figure}
The two conditions are given by
\begin{align}\label{eq:betacond}
\mu_l&= \mu_N-\mu_P = -\mu_I,\\
\label{eq:neutralitycond}
\frac{1}{2}n_B + n_I&=\sum_l n_l,
\end{align}
where $\mu_N$ and $\mu_P$ are the chemical potentials of the neutron and the proton, respectively , and $n_l$ is the lepton number density.
The leptonic free energy is that of a (massive) Fermi gas:
\begin{equation}\label{eq:Omegalepton}
\Omega_\ell = -\frac{1}{24\pi^2}\Theta\left(\mu_\ell-m_\ell\right)\left[\left(2\mu_\ell^2-5m_\ell^2\right)\mu_\ell\sqrt{\mu_\ell^2-m_\ell^2}+3m_\ell^4\log\left(\frac{\sqrt{\mu_\ell^4-m_\ell^4}+\mu_\ell}{m_\ell}\right)\right],
\end{equation}
from which we can obtain the leptonic density as a function of the chemical potential:
\beq
n_\ell(\mu_\ell)=\Theta_H(\mu_\ell-m_\ell)\frac{(\mu_\ell^2-m_\ell^2)^{\frac32}}{3\pi^2}.
\label{eq:Fermi_rho}
\eeq
Plugging eqs.~\eqref{eq:Fermi_rho} and \eqref{eq:betacond} into eq.~\eqref{eq:neutralitycond} yields:
\beq\label{eq:betaeqfinal}
\frac{1}{2}n_B +n_I - \frac{|\mu_I|^3}{3\pi^2}\left[   \Theta_H(-\mu_I-m_\mu)   \left(1-\frac{m_\mu^2}{\mu_I^2}\right)^{\frac{3}{2}}  +\Theta_H (-\mu_I) \right]=0,
\eeq
where we have approximated the electron to be massless. The isospin charge density is obtained from the near-boundary behavior of $a'_0(z)$, which is determined by the parameters $n_B$, $\mu_I$, so that solving  eq.~\eqref{eq:betaeqfinal} provides $\mu_I(n_B)$.

Having now a family of solutions determined by the single parameter\footnote{Note, however, that eq.~\eqref{eq:betaeqfinal} is derived assuming that the parameters $n_B$ and $n_I$ are read off of the asymptotics of the $A_t$ field rescaled with the overall coefficient of the action, as per standard holographic dictionary: This implies that for matching procedure (a), they both get rescaled with $c_b$. 
Because of this, the equation of state will now depend on $c_b$ in a nontrivial way: Before the introduction of $\beta$-equilibrium, it amounts to an overall rescaling of density, energy and pressure. We are considering now to have already fixed $c_b$, $b_1$ to the values appropriate for the matching, while practically the procedure involves solving eq.~\eqref{eq:betaeqfinal} at a fixed matching density for many values of $(b_1,c_b)$ and determining the pairs that produce a continuous equation of state after matching.} $n_B$, we can proceed to compute the equation of state.

To do so we compute the energy density $\mathcal{E}$ and pressure $P$ of the system. For a homogeneous system, the pressure is given in terms of the free energy $\Omega$ as $P=-\Omega$. For the leptons, the free energies $\Omega_e,\Omega_\mu$ are given by eq~\eqref{eq:Omegalepton}, while for the baryonic matter we can employ the holographic map to identify the free energy $\Omega_b$ with:
\beq
\Omega_b=-\left(S_{\rm TDBI}^{\rm on-shell}+S_{\rm TCS}^{\rm on-shell}\right). 
\eeq
The total pressure is then given by:
\beq
P= -\Omega_b-\Omega_e-\Omega_\mu ,
\eeq
while the total energy density is obtained by accounting for the chemical potentials and number densities as:
\beq
\mathcal{E}=-P +\mu_B n_B + \mu_I n_I + \mu_e n_e+\mu_\mu n_\mu .
\eeq
While the calculation up to now is sufficient for computing the EOS for the baryonic phase above the transition (matching) density $n_t$, it does not account yet for the possibility of a quark matter phase, which will be discussed in the next section.

Another quantity that can be computed from the solution of the $\beta$-equilibrium is the proton fraction (and the fractions of other particles) as a function of density, see fig.~\ref{fig:proton_fraction}. As a direct consequence of the smaller (and more realistic) symmetry energy of the EOS obtained via matching procedure (a), the corresponding proton fractions are also lower, and the gap between them and realistic ones (in the plot obtained from the SLy4 EOS) gets smaller for softer hybrid EOS. This trend is consistent with the results we will obtain for mass-radius (MR) curves of neutron stars: Matching the V-QCD-derived EOS with softer ones produces more phenomenologically acceptable MR curves when the matching is performed at transition densities $n_t$ in the range $n_t\in[1.2 n_S,1.8 n_S]$.

\section{Quark phase}\label{sec:quark_phase}

The onset of quark matter can be a decisive factor in establishing the highest mass a neutron star can achieve. If we want to compare our results with observations, we need to account for the possibility of quark matter influencing the MR curves of neutron stars.
To model the dense quark-matter phase, we follow~\cite{Alho:2012mh,Alho:2013hsa} and compute the equation of state from the thermodynamics of charged black hole solutions in V-QCD. For this computation, we
need to include the effects of backreaction of the gauge fields onto the geometry: To turn on a nontopological charge density, a horizon in the background is needed to impose Dirichlet boundary conditions on the Abelian vectorial gauge field $\widehat{A}_t$. Other components of the gauge field are set to zero.
For the metric, we take the Ansatz of eq.~\eqref{eq:metric} but instead of setting the blackening factor $f(r)$ to one it is now a free function.
We find  the solution that describes a geometry with a planar horizon ($f(r_h)=0$ for some value of $r=r_h$) by numerically solving the equations of motion arising from the gluon and flavor actions in eqs.~\eqref{eq:Sgdef} and~\eqref{eq:TDBI}. 
For the potentials we are using, only a ``hairless'' chirally symmetric black hole phase solution appears, so that the tachyon field $T$ vanishes. This solution is dual to a chirally symmetric quark-gluon plasma.

At zero quark masses, and accounting for the presence of the strange quark, $\beta$-equilibrium is trivially solved by an isospin symmetric configuration $n_u=n_d=n_s$, setting $\mu_I=0$ and similarly having vanishing lepton number densities. This is of course an approximation, in reality we should account for quark masses, including the larger strange mass.
The onset of the quark phase is then uniquely determined by requiring continuity of the pressure and the baryon number chemical potential $\mu_B$ at the transition, while the baryon number density is in general discontinuous. We follow the approach~\cite{Ishii:2019gta,Ecker:2019xrw,Jokela:2020piw} where the pressure from nuclear matter is directly compared to the pressure of quark matter (or, more precisely, the pressure difference between the quark matter and the vacuum) despite the fact that the nuclear matter pressure was computed in the probe approximation whereas the quark matter solution is fully backreacted.

\begin{figure}[!ht]
  \centering
  \stackon[5pt]{\includegraphics[width=0.49\textwidth]{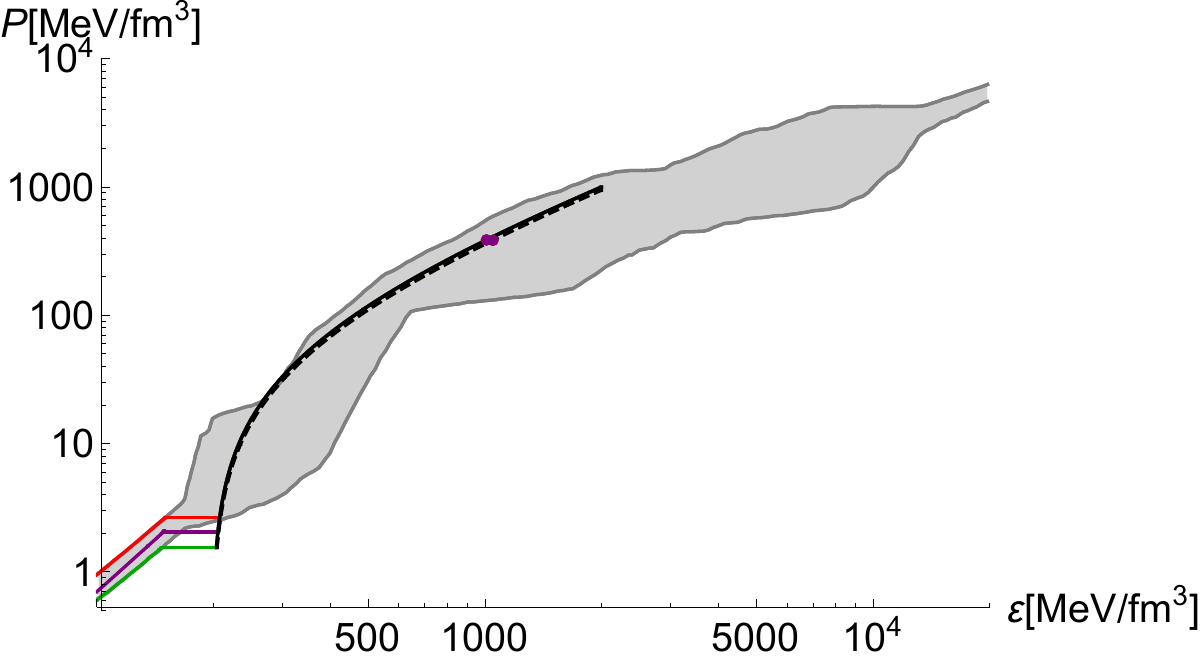}}{\footnotesize Procedure (b)}
  \stackon[5pt]{\includegraphics[width=0.49\textwidth]{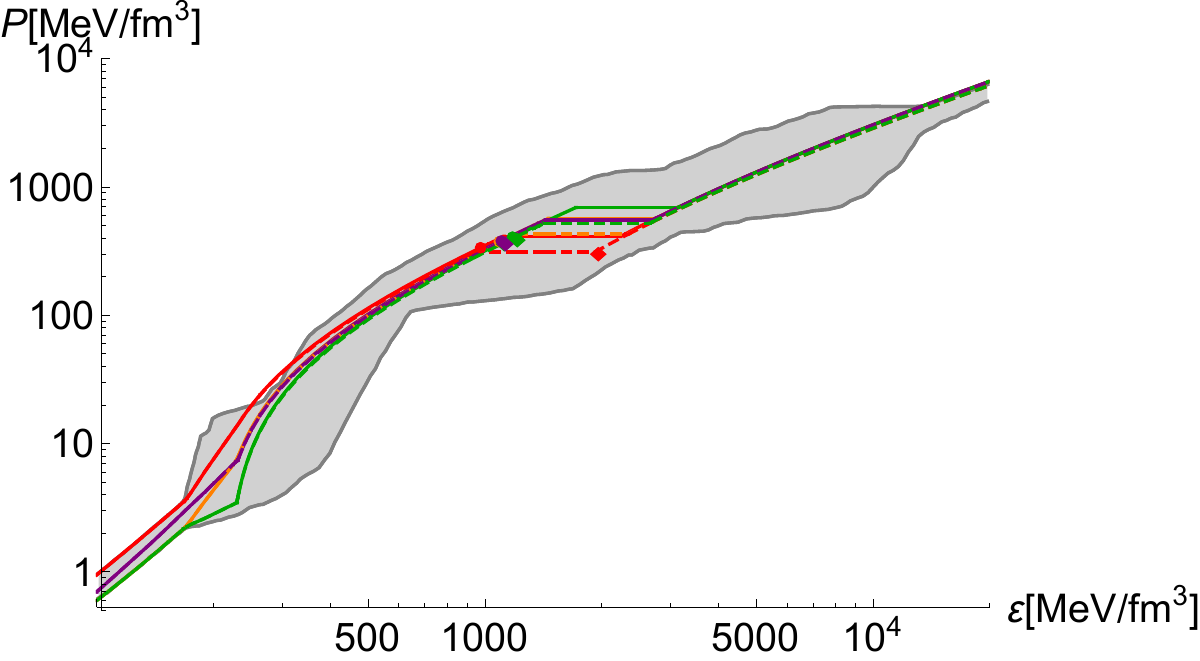}}{\footnotesize Procedure (a) $n_t=1.5n_S$}\\[10pt]
  \stackon[5pt]{\includegraphics[width=0.49\textwidth]{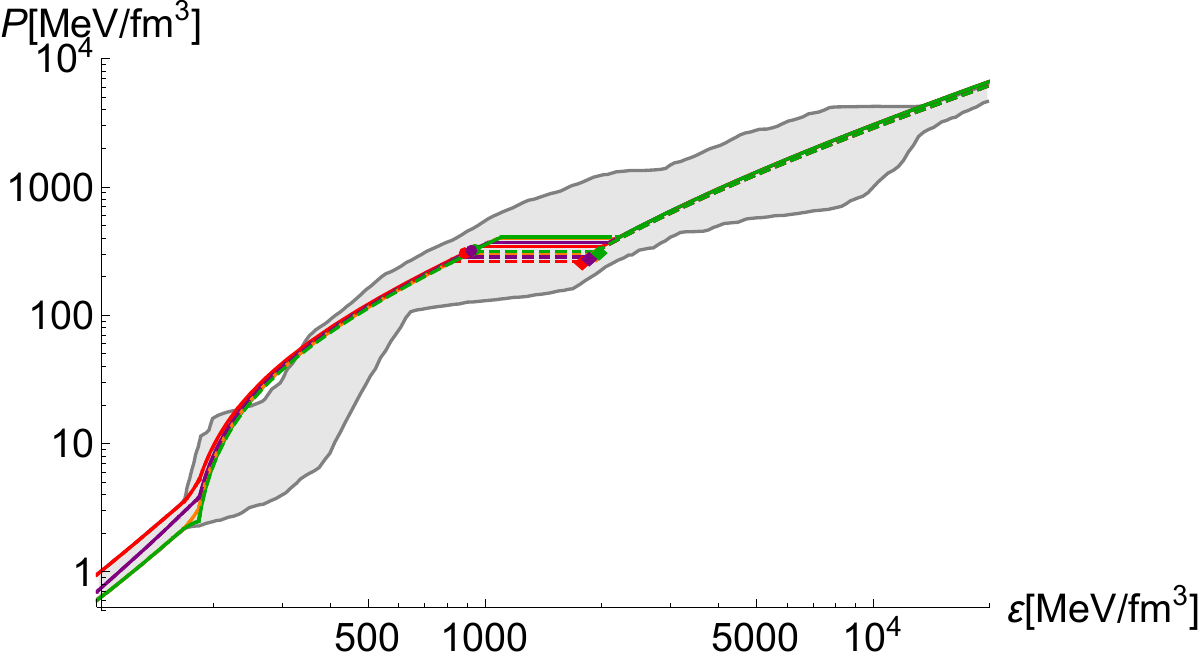}}{\footnotesize Procedure (a) $n_t=1.2n_S$}
  \stackon[5pt]{\includegraphics[width=0.49\textwidth]{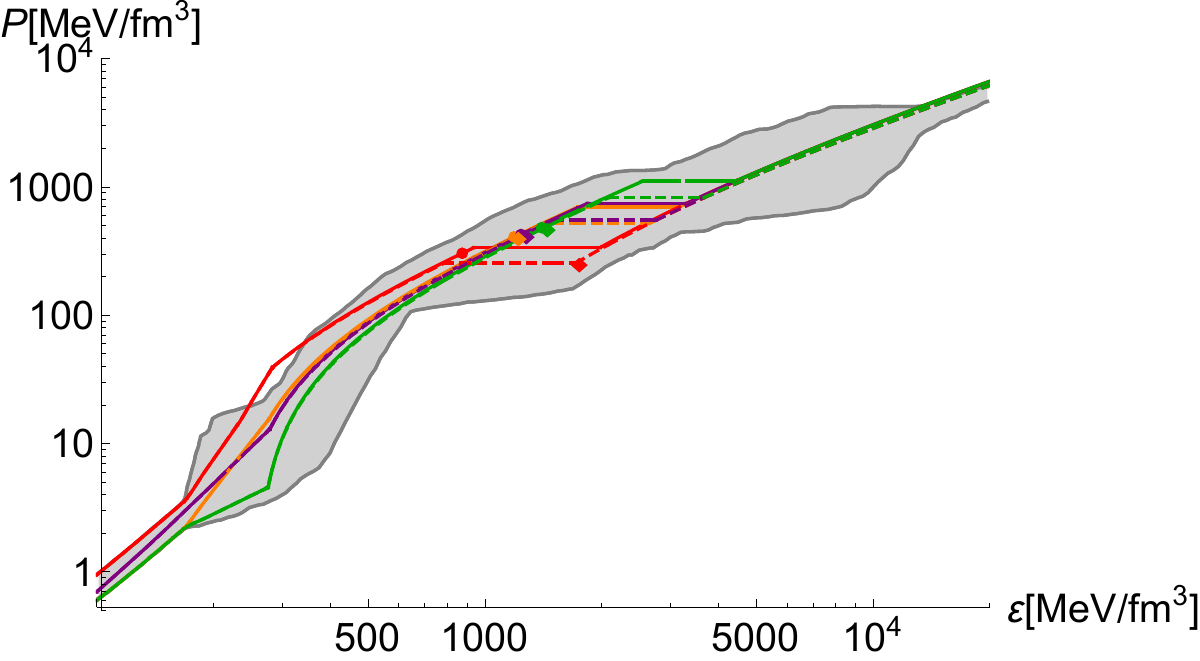}}{\footnotesize Procedure (a) $n_t=1.8n_S$}
  \caption{Collection of equations of state. Color coding and line styles are as in fig.~\ref{fig:esymmetry}: Green for HLPS soft, orange for HLPS intermediate, red for HLPS stiff, and purple for SLy4; solid (dashed) lines correspond to potential 7a (5b). The top-left panel corresponds to procedure (b), with the black segments representing the V-QCD high-density part. The other panels correspond to procedure (a) with transition densities $n_t = \{1.2,1.5,1.8\}n_S$. Colored dots (diamonds) mark the central energy densities and pressures of the most massive neutron stars supported by each EOS. The gray band is obtained from all the quadrutropic EOS interpolating from EFT results (low energy density) to perturbative QCD ones (high energy density) by requiring the TOV mass $M_{\rm TOV}$ and the tidal deformability $\Lambda$ to satisfy $M_{\rm TOV}>2 M_{\odot}$  and $\Lambda_{1.4 M_\odot}<580$ \cite{Jarvinen:2021jbd}.}
  \label{fig:EOS}
  \end{figure}

In fig.~\ref{fig:EOS} we present the hybrid equations of state built with the methodology described for each combination of matching procedure, potential, low density behavior and transition density, for a total of 30 equations.

\begin{figure}[!ht]
  \centering
  \stackon[5pt]{\includegraphics[width=0.49\textwidth]{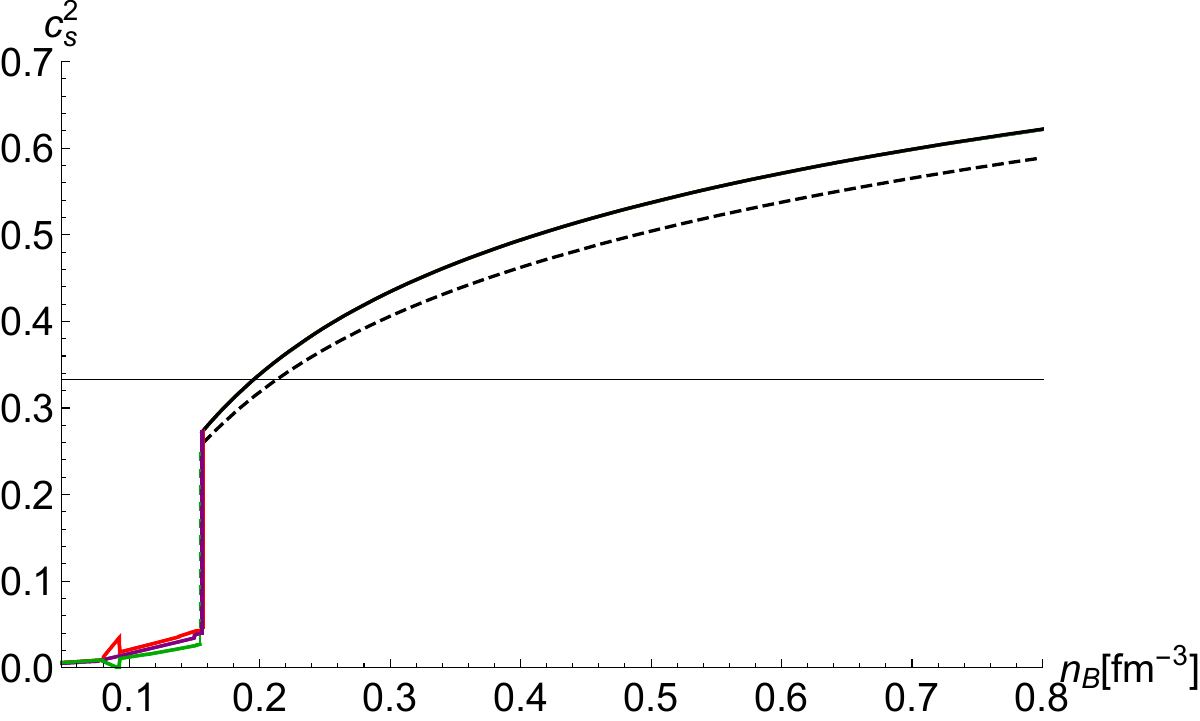}}{\footnotesize Procedure (b)}
  \stackon[5pt]{\includegraphics[width=0.49\textwidth]{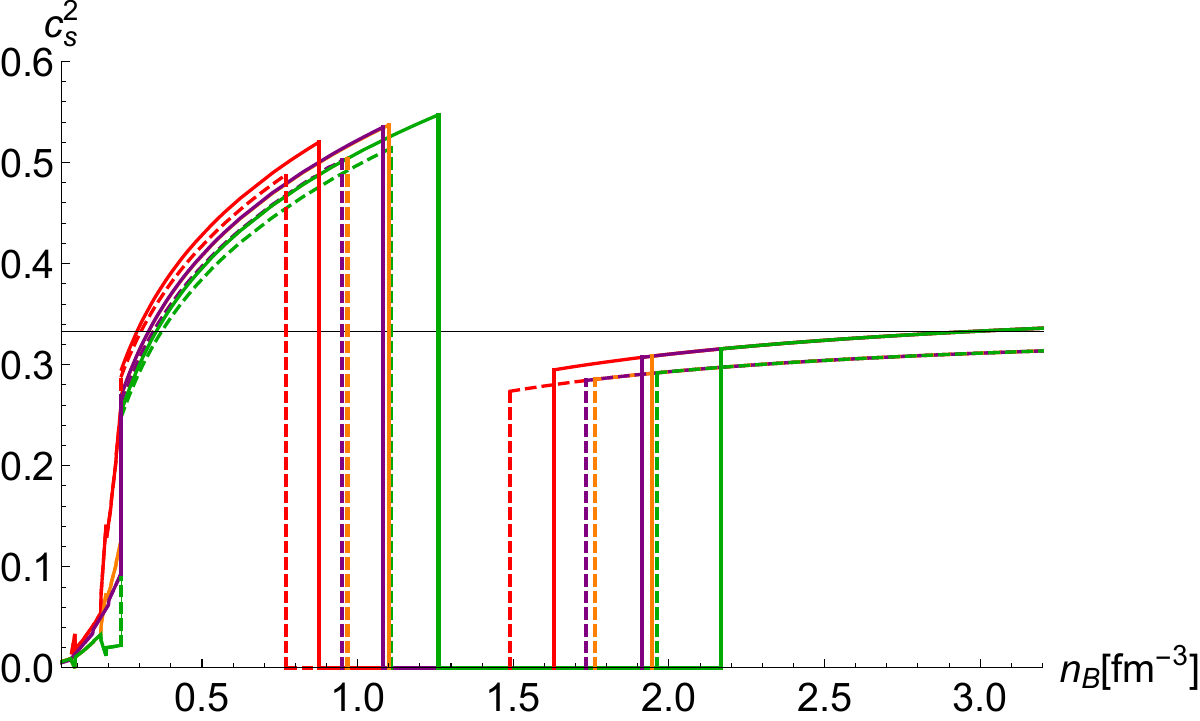}}{\footnotesize Procedure (a) $n_t=1.5n_S$}\\[10pt]
  \stackon[5pt]{\includegraphics[width=0.49\textwidth]{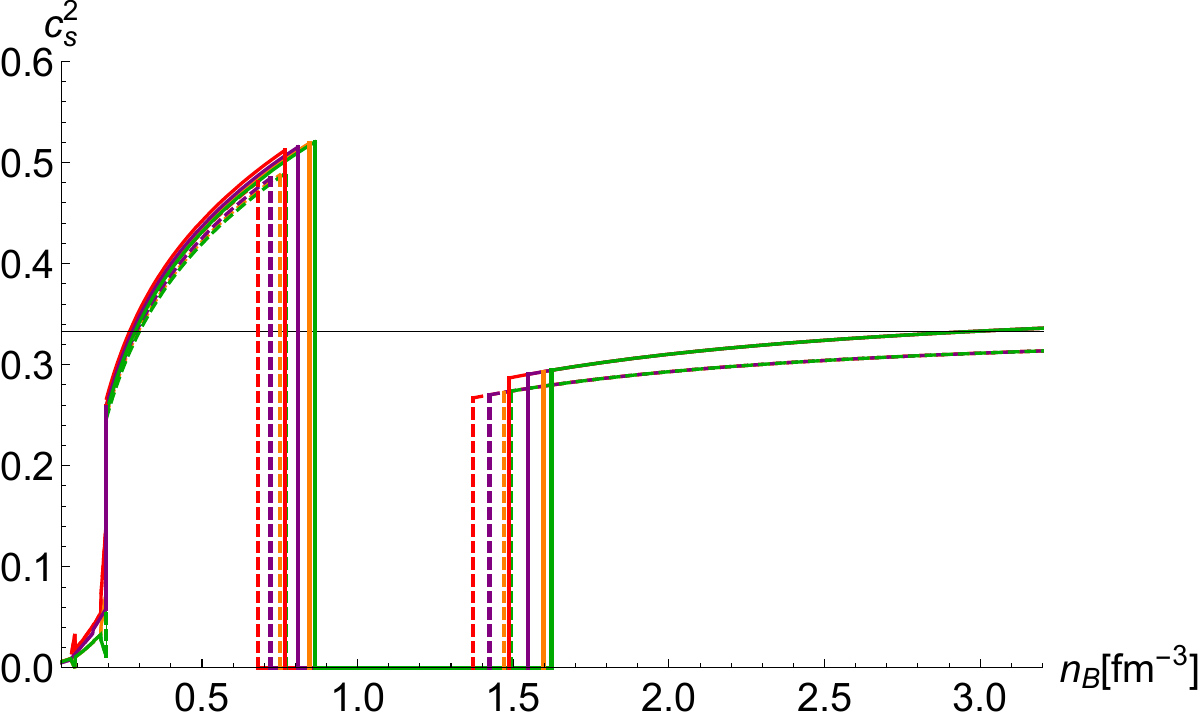}}{\footnotesize Procedure (a) $n_t=1.2n_S$}
  \stackon[5pt]{\includegraphics[width=0.49\textwidth]{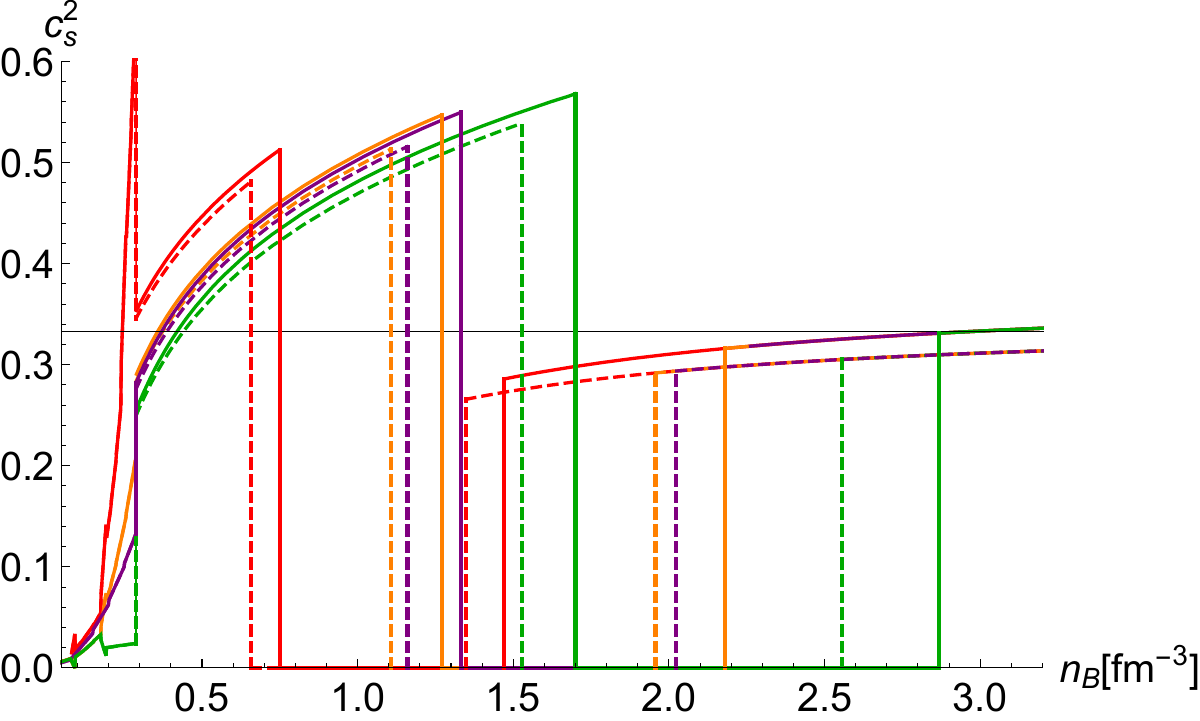}}{\footnotesize Procedure (a) $n_t=1.8n_S$}
  \caption{Speed of sound squared as a function of the baryon number density for each hybrid EOS. The black horizontal line corresponds to the conformal limit $c_s^2 = 1/3$. Color coding and line styles are as in fig.~\ref{fig:esymmetry}: Green for HLPS soft, orange for HLPS intermediate, red for HLPS stiff, and purple for SLy4; solid (dashed) lines correspond to potential 7a (5b). The top-left panel refers to matching procedure (b), while the others correspond to procedure (a) with transition densities $n_t = \{1.2,1.5,1.8\}n_S$. Horizontal gaps indicate the first-order phase transition between baryonic and quark matter phases, bounded by vertical guide lines.}
  \label{fig:csvsd}
  \end{figure}
With the equations of state of fig.~\ref{fig:EOS} in hand, we can compute the speed of sound squared numerically as
\beq
c_s^2=\frac{\p P}{\p\mathcal{E}}.
\eeq
In fig.~\ref{fig:csvsd}, we display the speed of sound squared for all the EOSs computed, but take into account also the transition to the quark matter phase, which manifests itself as a horizontal gap in the plots.
Note that we could only take the quark matter phase transition into account for the procedure (a).

\section{Neutron stars}\label{sec:neutron_stars}

We want to compute the properties of neutron stars resulting from the set of equations of state that we have built: To do so, we solve the Tolman-Oppenheimer-Volkov (TOV) equations, that describe the coupling between a static, spherically symmetric distribution of matter and gravity. The set of coupled equations is given by
\begin{align}\label{eq:TOV1}
  \frac{\d P}{\d r}&= -G(\calE+ P)\frac{m+4\pi r^3 P}{r(r-2Gm)},\\
  \frac{\d m}{\d r}&= 4\pi r^2 \calE.\label{eq:TOV2}
\end{align}
The TOV equations are solved with initial values $P(r=0)=P_0$, $m(r=0)=0$ and the value $R$ is determined from $P(R)=0$, which is identified as the radius of the star (correspondingly, the mass of the star is given by $M=m(R)$). Repeating the process for all values of $P_0$ that result in stable stars leads to  
a curve in the $M$-$R$ plane describing all possible neutron stars for each given EOS.

\begin{figure}[!ht]
  \centering
  \stackon[5pt]{\includegraphics[width=0.49\textwidth]{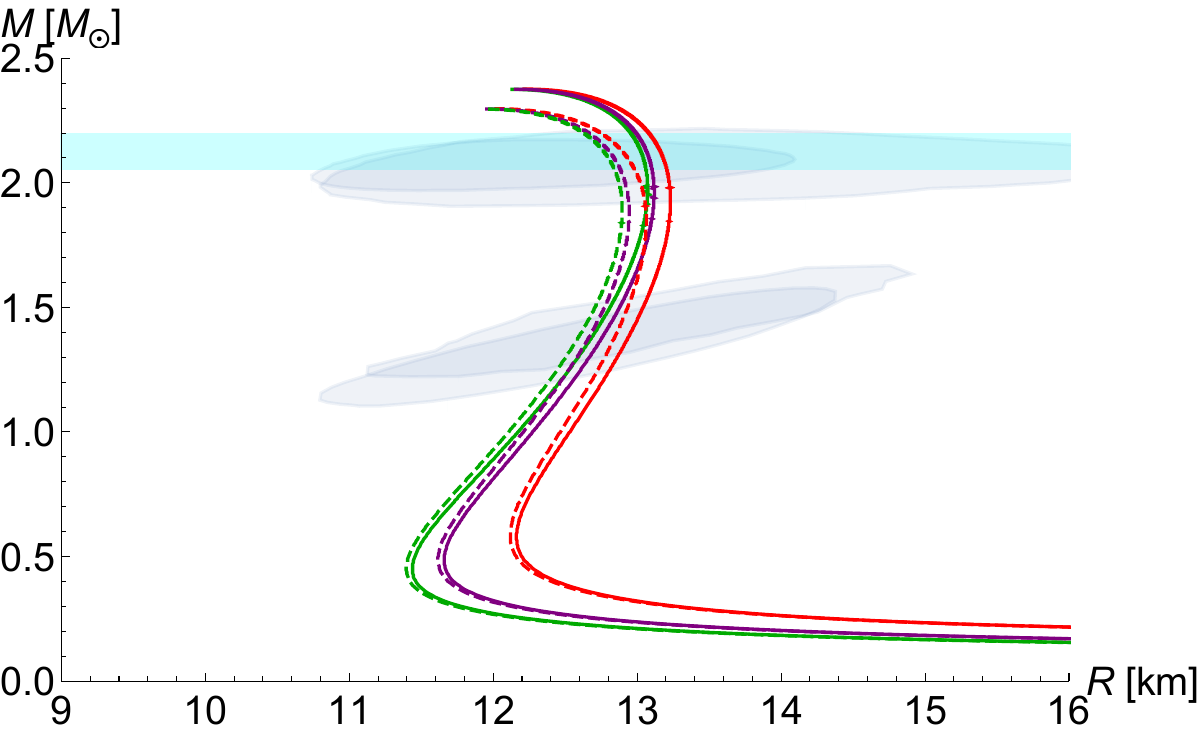}}{\footnotesize Procedure (b)}
  \stackon[5pt]{\includegraphics[width=0.49\textwidth]{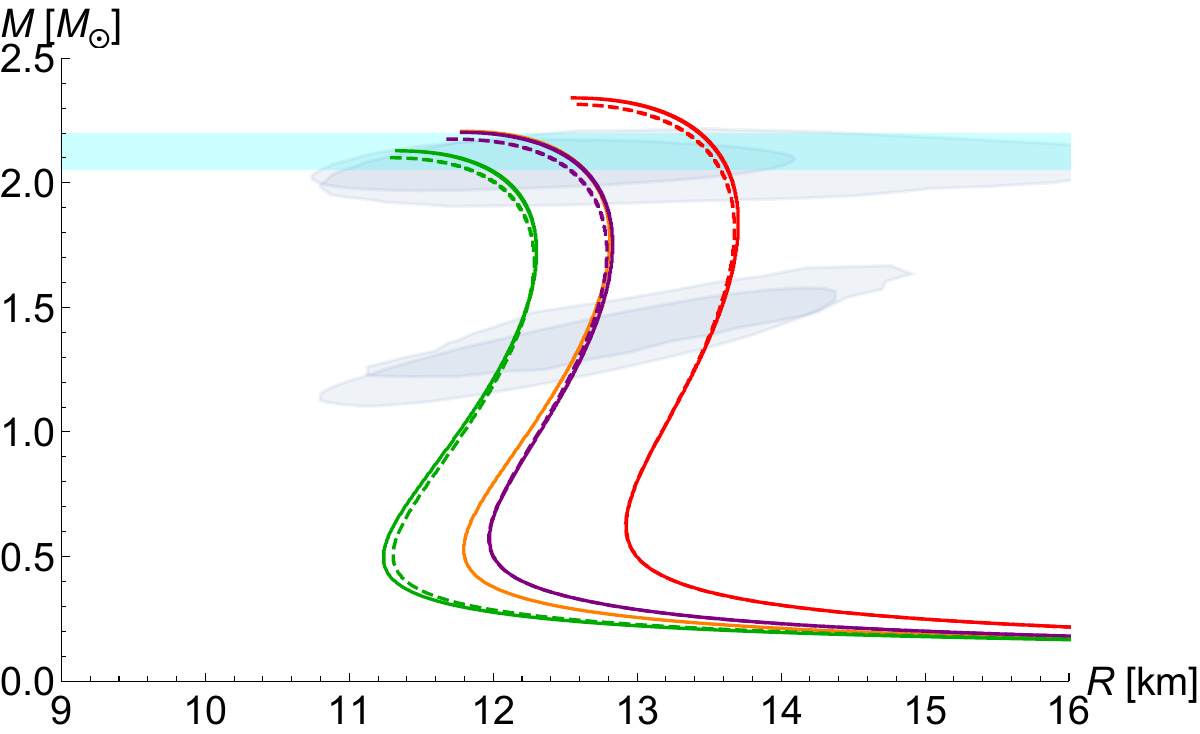}}{\footnotesize Procedure (a) $n_t=1.5n_S$}\\[10pt]
  \stackon[5pt]{\includegraphics[width=0.49\textwidth]{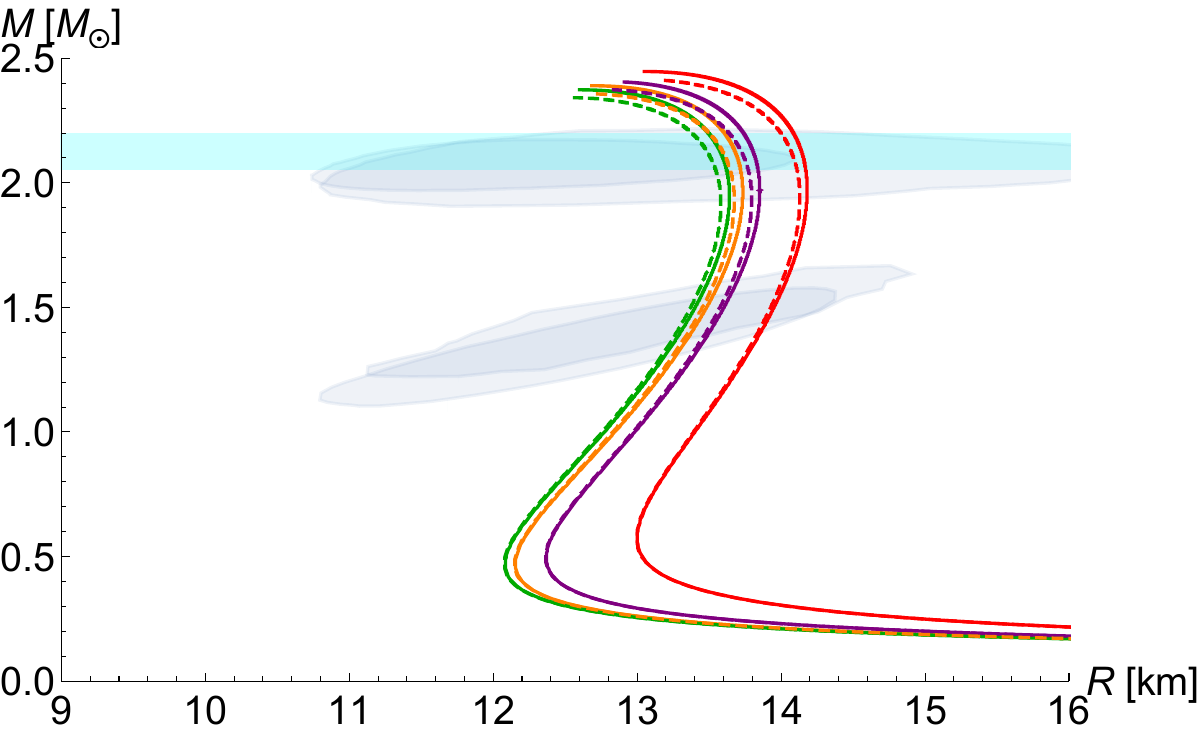}}{\footnotesize Procedure (a) $n_t=1.2n_S$}
  \stackon[5pt]{\includegraphics[width=0.49\textwidth]{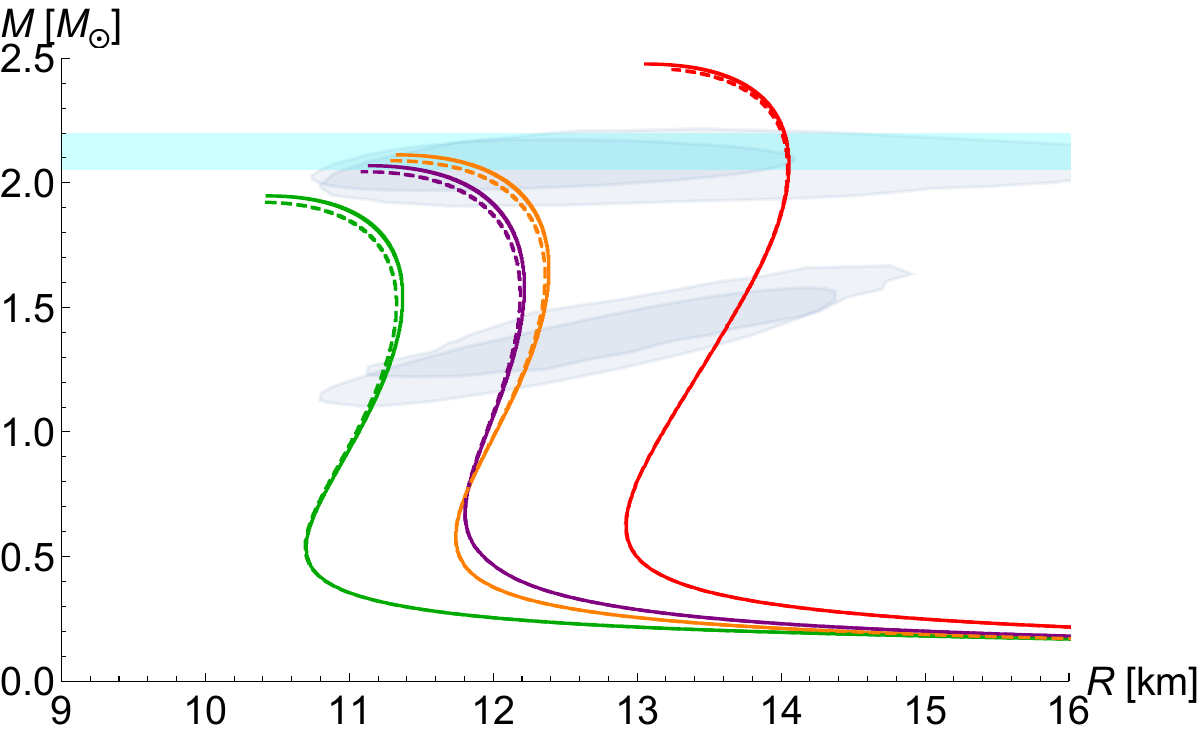}}{\footnotesize Procedure (a) $n_t=1.8n_S$}
  \caption{Mass-radius (MR) relations for neutron stars obtained by solving the TOV equations with the constructed EOS. Each curve represents a different hybrid EOS, with color coding and line styles as in fig.~\ref{fig:esymmetry}: Green for HLPS soft, orange for HLPS intermediate, red for HLPS stiff, and purple for SLy4; solid (dashed) lines correspond to potential 7a (5b). The top-left panel corresponds to matching procedure (b), while the others show procedure (a) with $n_t = \{1.2,1.5,1.8\}n_S$. The gray shaded region marks observational constraints from NICER, and the blue shaded area indicates an estimate for the highest possible mass value $M_{\rm TOV}$ for a static neutron star, obtained combining results from NICER and the analysis of GW170817.}
  \label{fig:MR}
  \end{figure}

Our results for this procedure are shown in fig.~\ref{fig:MR}, where we also report the constraints from NICER~\cite{Riley:2019yda,Miller:2019cac,Riley:2021pdl,Miller:2021qha} and an estimate on the highest possible mass as determined in ref.~\cite{Margalit:2017dij}. Almost all of our EOSs are stiff enough to reach the range of two solar masses as required, for example, by the NICER measurement of the mass of J0740+6620. However, the EOSs with $n_t=1.8n_S$ and the soft variant of the HLPS EOSs are so soft that they are in tension with this measurement. Moreover, many other EOSs, including all EOSs constructed with the procedure (b), lead to slightly larger maximum masses than the bound of~\cite{Margalit:2017dij}. Note however that this bound is a rather indirect result of the  analysis of the GW170817 binary merger event.

Note that none of our equations of state can support stable quark matter cores in neutron stars~\cite{Jokela:2020piw}. However, the detailed picture is different depending on the choice potentials in the V-QCD action.
For all of our hybrid EOSs using potentials 7a, as the mass is increased, the star becomes unstable even before the phase transition is reached in the core of the star, while for potentials 5b the phase transition happens at a lower energy density, so that the transition is reached in the cores of the heaviest neutron stars constructed from six equations of state.

\begin{figure}[!ht]
  \centering
  \stackon[5pt]{\includegraphics[width=0.49\textwidth]{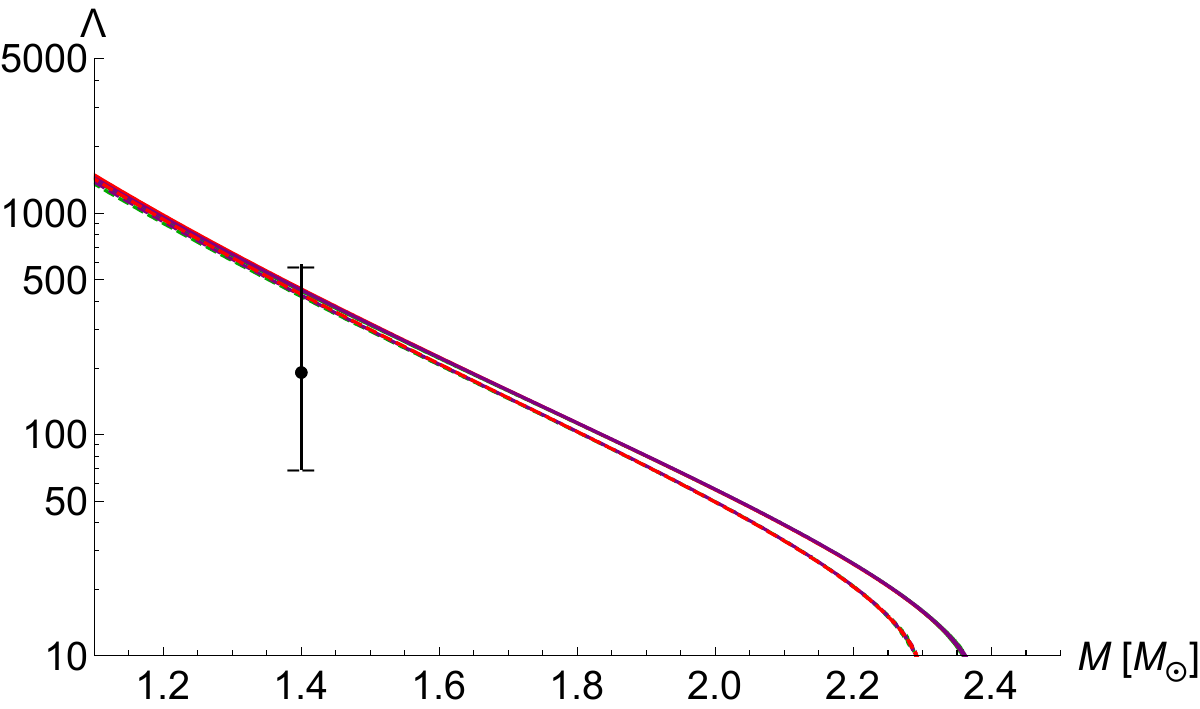}}{\footnotesize Procedure (b)}
  \stackon[5pt]{\includegraphics[width=0.49\textwidth]{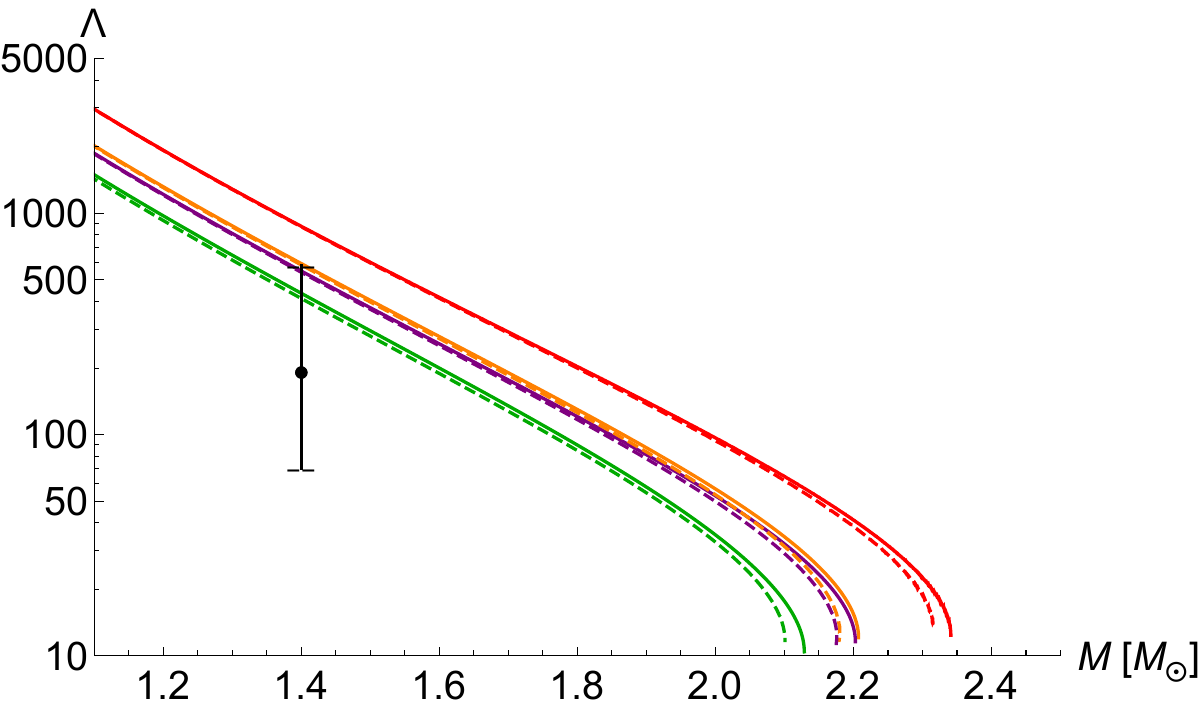}}{\footnotesize Procedure (a) $n_t=1.5n_S$}\\[10pt]
  \stackon[5pt]{\includegraphics[width=0.49\textwidth]{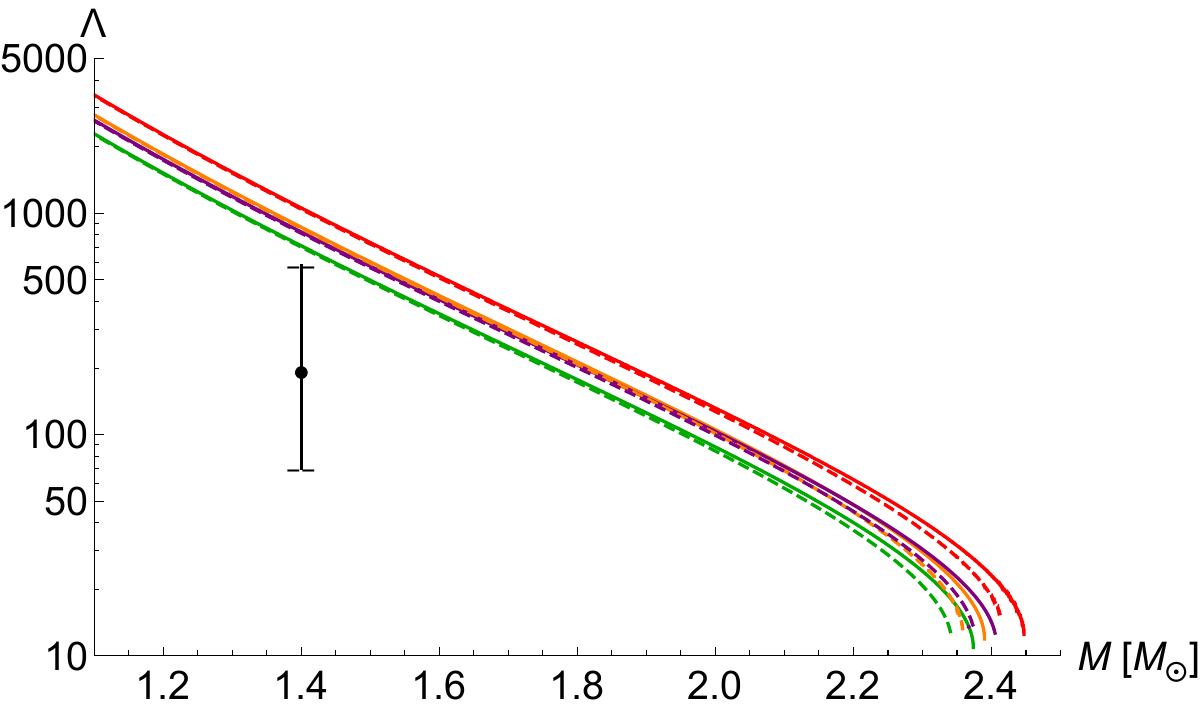}}{\footnotesize Procedure (a) $n_t=1.2n_S$}
  \stackon[5pt]{\includegraphics[width=0.49\textwidth]{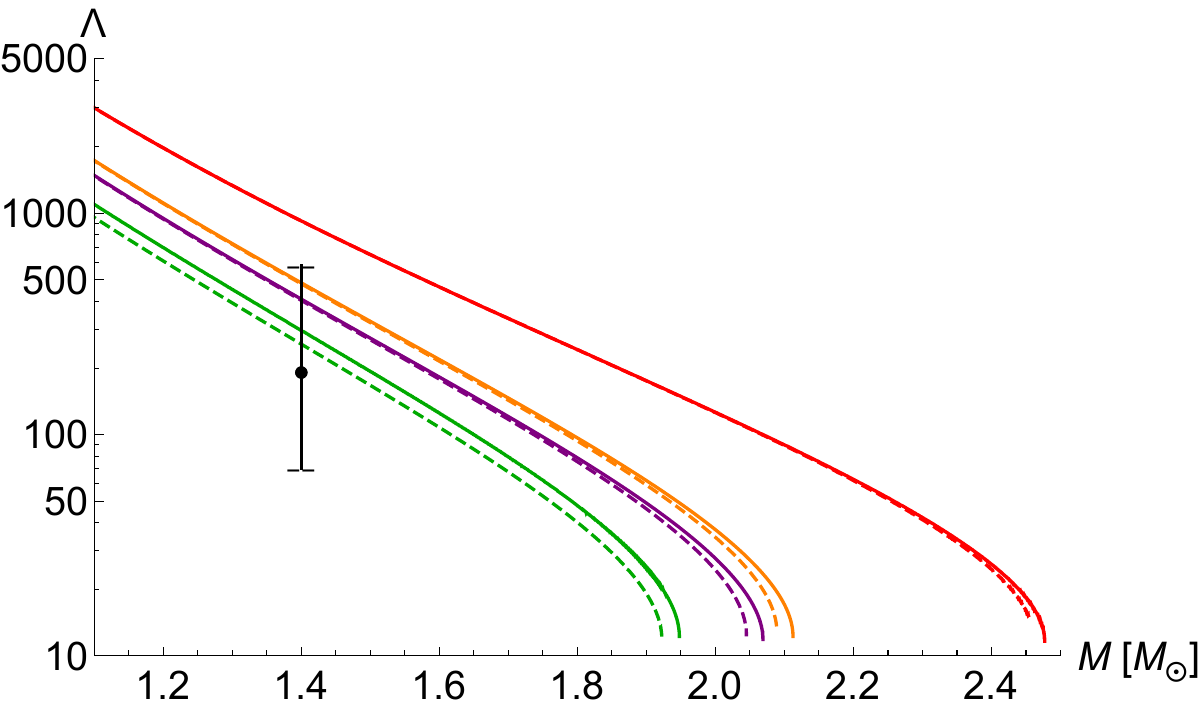}}{\footnotesize Procedure (a) $n_t=1.8n_S$}
  \caption{Mass-Tidal Deformability (MTD) curves for neutron stars. Each curve represents a different hybrid EOS, with color coding and line styles as in fig.~\ref{fig:esymmetry}: Green for HLPS soft, orange for HLPS intermediate, red for HLPS stiff, and purple for SLy4; solid (dashed) lines correspond to potential 7a (5b). The top-left panel corresponds to matching procedure (b), while the others show procedure (a) with $n_t = \{1.2,1.5,1.8\}n_S$.}
  \label{fig:MTidal}
  \end{figure}

Another property that we can compute having access to the EOS, is the tidal deformability $\Lambda$: It is defined from the tidal Love number $k_2$ and the star compactness $c=\frac{GM}{R}$ as 
\beq
\Lambda=\frac{2k_2}{3c^2}.
\eeq
In fig.~\ref{fig:MTidal} is shown the tidal deformability as function of the mass obtained for each EOS (MTD curves), together with the current most stringent bound from LIGO/Virgo GW\-170817 for a neutron star of mass $1.4 M_\odot$. We see that models using the stiff variant of the HLPS EOS are in slight tension with the tidal deformability measurement.

\begin{figure}[!ht]
  \centering
  \stackon[5pt]{\includegraphics[width=0.49\textwidth]{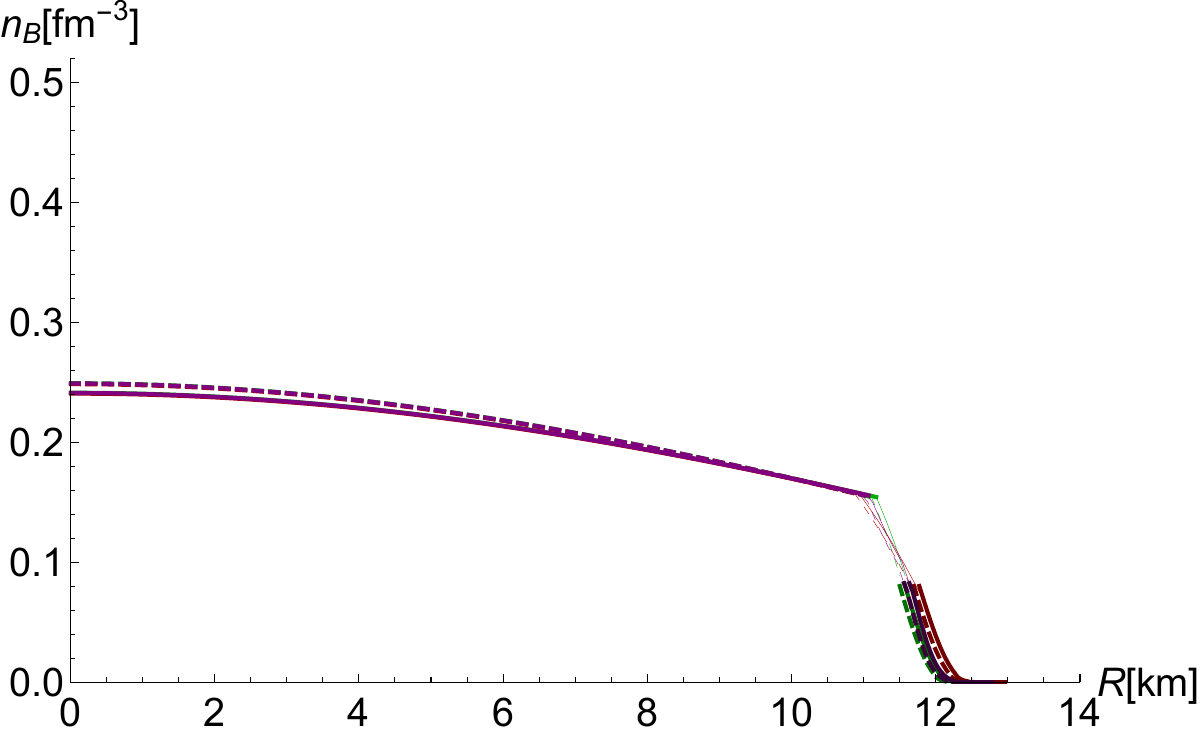}}{\footnotesize Procedure (b)}
  \stackon[5pt]{\includegraphics[width=0.49\textwidth]{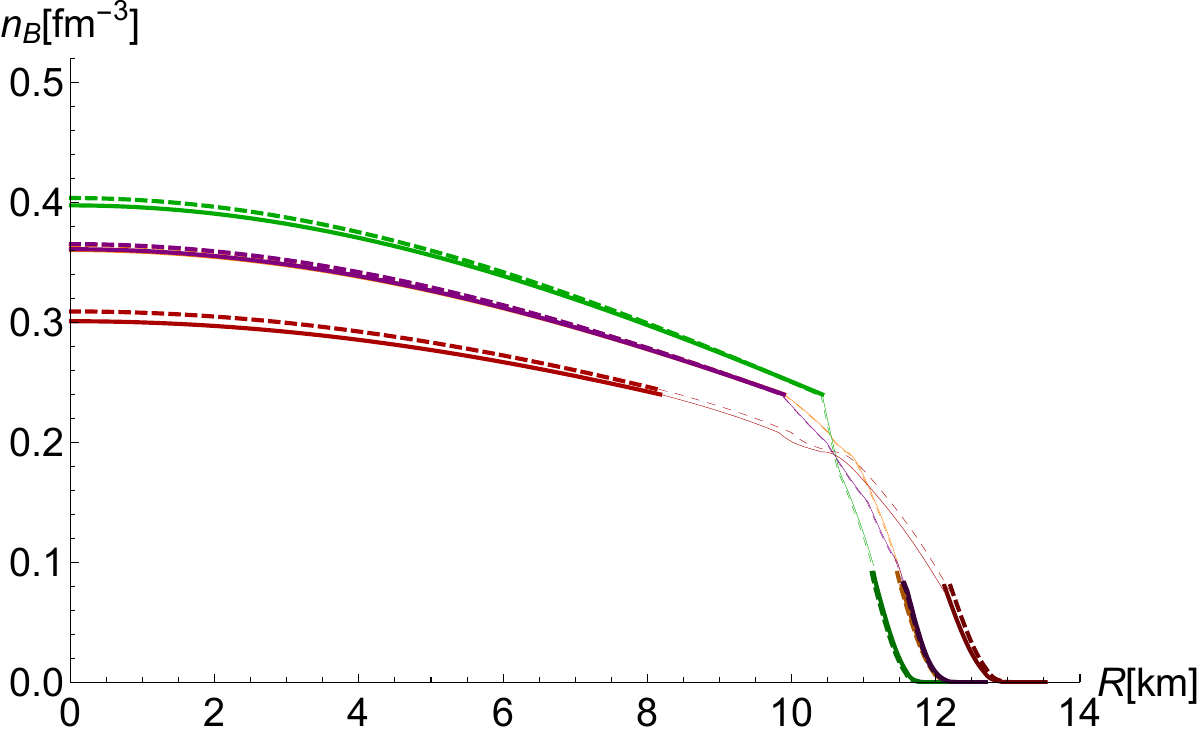}}{\footnotesize Procedure (a) $n_t=1.5n_S$}\\[10pt]
  \stackon[5pt]{\includegraphics[width=0.49\textwidth]{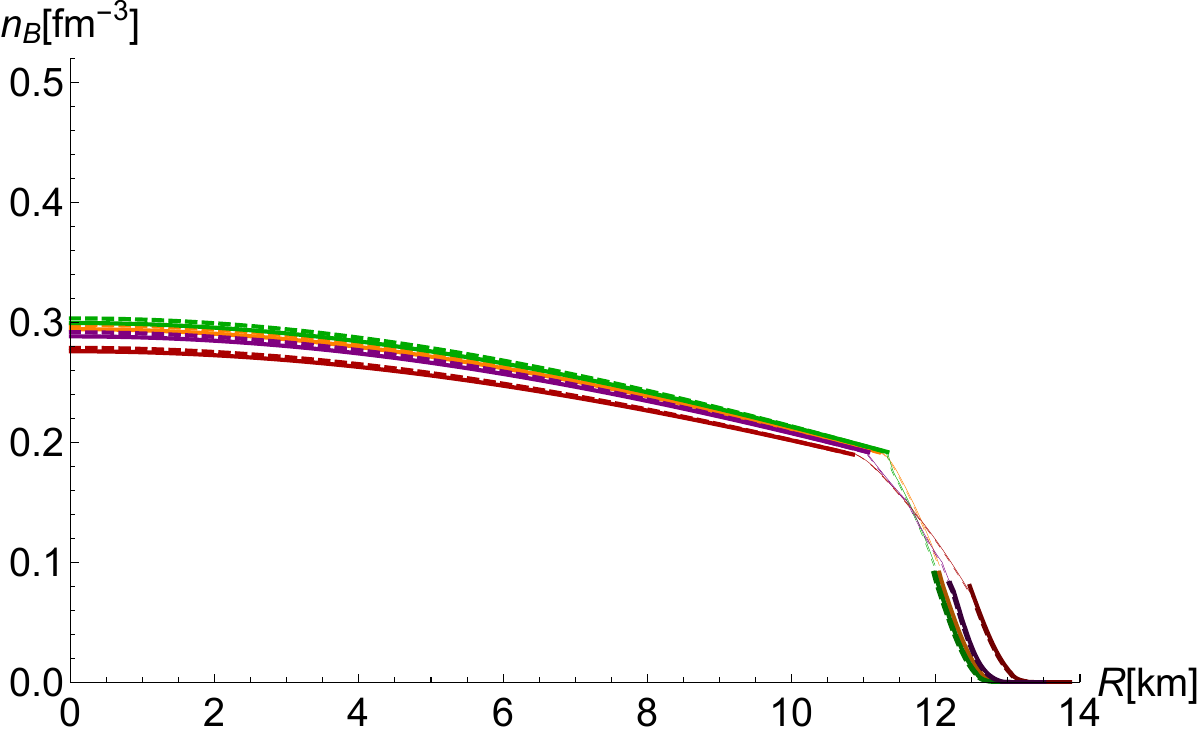}}{\footnotesize Procedure (a) $n_t=1.2n_S$}
  \stackon[5pt]{\includegraphics[width=0.49\textwidth]{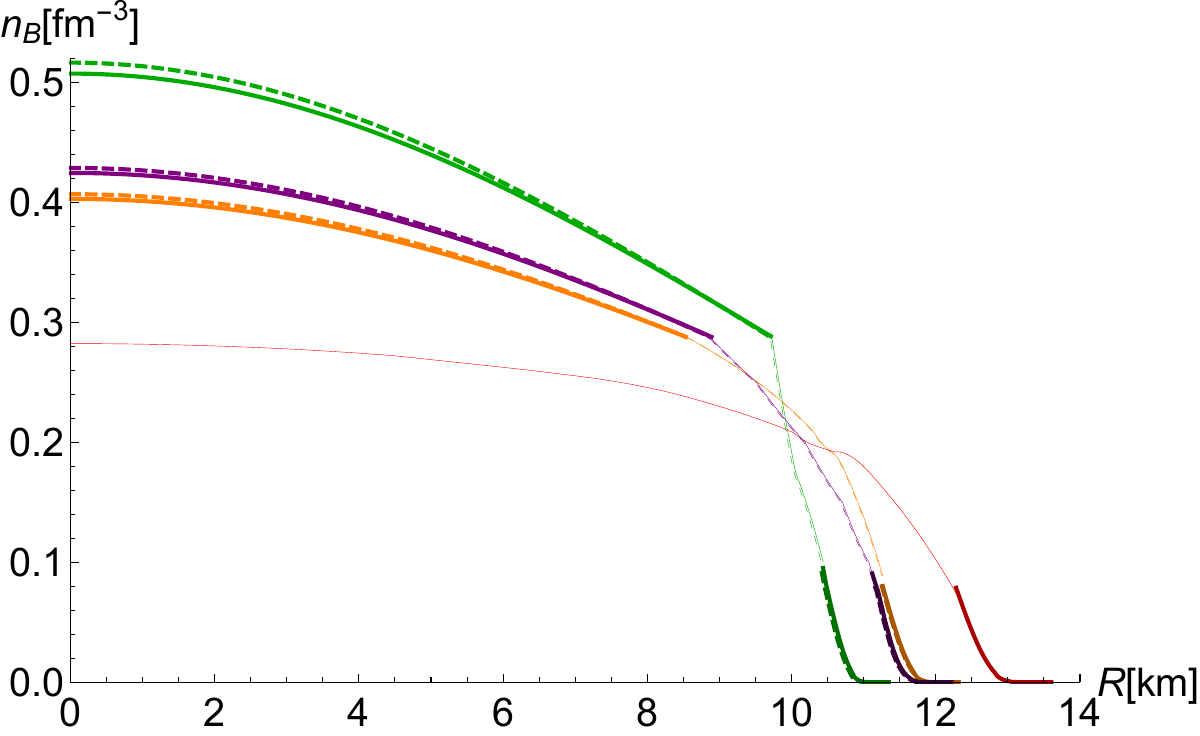}}{\footnotesize Procedure (a) $n_t=1.8n_S$}
  \caption{Density as a function of radial distance from the center of neutron stars, for typical stars of mass $1.4M_\odot$. Top left panel: Matching procedure (b). Remaining panels: Matching procedure (a) with matching density $n_t =\{1.2,1.5,1.8\}n_S$. The color coding and line styles are as in fig.~\ref{fig:esymmetry}: Green for HLPS soft, orange for HLPS intermediate, red for HLPS stiff, and purple for SLy4; solid (dashed) lines correspond to potential 7a (5b). Thick lines, thin lines, and darker thick lines correspond respectively to the holographic core, the low density core (HPLS or SLy4) and the crust of the stars. Note that in the bottom-right panel, only one red curve appears; this is because for this star the density never reaches the matching density $n_t=0.288 \rm{fm}^{-3}$, and so it is insensitive to the holographic model.}
    \label{fig:dR}
      \end{figure}
Finally, in fig.~\ref{fig:dR} we display for illustrative purposes the density and composition of the entire neutron star as a function of its radius.
In particular, the thick lines illustrate how large a fraction of the neutron star is made of the holographic QCD matter and what is made by the low-energy phenomenologically established equations of states.
As an interesting example to point out, for the low matching density of $n_t=1.2n_S$ (bottom-left panel) the holographic EOS is carrying most of the core of the star and does not change much with respect to the different low-energy EOSs (all about 11 to 11.5 km). On the other hand, for the high matching density $n_t=1.8n_S$ (bottom-right panel), the holographic EOS simply does not enter the hybrid star that is soft in the low-density region (red thin line in the figure). 

\section{Conclusions}\label{sec:conclusions}

With this work, we have studied the effects of isospin breaking in nuclear matter within the holographic V-QCD model. 
A crucial simplifying approximation utilized in this work is the homogeneous Ansatz for nuclear matter, which consists of introducing a finite baryonic density via a discontinuity in the bulk gauge fields, which are taken to be independent of the position in three dimensional space. 
Isospin is introduced via an external chemical potential, following the usual holographic prescription of encoding its value in the UV boundary condition for the field $A^{a=3}_t$.
The approximation inherent in the homogeneous Ansatz is expected to be reliable only at high densities: For this reason, we used a set of phenomenological equations of state for the description of the low-density regime (SLy4 and three representative HLPS). We considered two different approaches to matching the low density EOS with the holographic high density ones: The one labeled ``procedure (a)'' is more phenomenology-driven and corresponds to the introduction of a new parameter that acts as a rescaling factor for the holographic action, while with the one labeled ``procedure (b)'' avoid introducing new parameters and instead fit the already present ones to correctly reproduce the value of saturation density.

With this setup, expanding around the isospin symmetric configuration, we computed the symmetry energy of nuclear matter as a function of density, and compared its value around saturation density with phenomenology: We found remarkable agreement for procedure (a), while procedure (b) overestimates the symmetry energy at saturation, a result shared (though with milder magnitude) with other holographic models. Relaxing the assumption of small isospin, we solved the full system at arbitrary isospin and computed the EOS of $\beta$-equilibrated nuclear matter, complete with particle fractions (we have included massless electrons and massive muons to impose electric charge neutrality). For matching procedure (a) we have also included a transition to quark matter, while for procedure (b) it turned out to be impossible because of the mismatch in the scales of baryonic chemical potentials introduced by the choice of the single fit parameter $b_1$: This problem is avoided in procedure (a) since the presence of two free parameters ($b_1$ and the rescaling $c_b$) allows for a matching of the scales of both densities and chemical potentials. We interpret the need for more free parameters as the compound error accumulated from two approximations: The restriction to the homogeneous Ansatz without other phases for holographic nuclear matter, and the use of the flat space expression for the tachyon potential in the Chern-Simons term. It would be interesting to explore how the results would change by considering different functional forms for the dilaton potential, a task we leave for the future.

Finally, having the computed EOS in hand, we solved the TOV equations and derived the properties of static neutron stars. We computed MR curves and tidal deformabilities of the stars resulting from each EOS, finding results that are in all cases close to observations, and a subset of EOSs that satisfy the constraints from NICER on the MR curves and from LIGO/Virgo on the tidal deformability. In tab.~\ref{tab:parameters} we report the values of the parameters $n_t,b_1,c_b$ for each hybrid equation of state. On the left are the values for potentials 7a, while on the right for potentials 5b. In the last column labeled ``Data'' we evaluate the compatibility of the resulting neutron stars with the NICER and LIGO/Virgo data, indicating with a check mark \cmark\ full compatibility, while with \xmark\ the presence of tension with the data currently available. An interesting result is that those EOSs that produce neutron stars that satisfy these bounds are also those that correspond to parameters that correctly reproduce the phenomenological symmetry energy at saturation density: In particular, all of the hybrid EOSs built from the stiff HLPS low-density EOS are in tension with astrophysical bounds, and they all result in a high value for the symmetry energy at saturation $S_0=[36.4,39.4]$ MeV. For the ones built from the soft HLPS EOS we find that only one pair (corresponding to potential 5b and 7a) of hybrid EOSs is in good agreement with observational bounds, resulting from the matching at $n_t=1.5n_S$, for which $S_0=\{32.0,32.8\}$MeV. For the intermediate HLPS, the hybrid EOS pair that is most compatible with astrophysical bounds is that coming from the matching at $n_t=1.8n_S$, and the corresponding symmetry energy at saturation is $S_0=\{31.9,32.6\}$MeV. Finally, for the SLy4 EOS, both the hybrid pairs corresponding to $n_t={1.5,1.8}n_S$ are close to observational bounds: For $n_t=1.5n_S$ we have $S_0=\{33.7,34.3\}$MeV, while for $n_t=1.8n_S$ we find $S_0=\{31.3,32.0\}$MeV. This precise agreement with phenomenology may hint that the rescaling procedure (a) tends to successfully capture some physics that is lost when taking the approximations of homogeneity and of the flat-space dilaton potential.

\begin{table}[!ht]
		\centering
        \resizebox{\textwidth}{!}{
		\renewcommand{\arraystretch}{1.5} 
		\begin{tabular}{|c|c|c|c|c|}
			\hline
			\rowcolor{gray!20} \cellcolor{gray!70} \textcolor{white}{\textbf{V-QCD 7a}} & $n_t$ & $b_1$ & $c_b$ & Data\\
			\hline
			\cellcolor{gray!20} \textbf{HLPS soft} &  1.2 & 9.799 &  3.545 & \xmark\\
			\hline
			\cellcolor{gray!20} \textbf{HLPS int.} &  1.2 &  9.796 &  3.499 &\xmark\\
			\hline
			\cellcolor{gray!20} \textbf{HLPS stiff} & 1.2 &  9.867  &  3.287&\xmark \\
			\hline
			\cellcolor{gray!20} \textbf{SLy4} & 1.2 & 9.86 &  3.402 & \xmark\\
			\hline
			\cellcolor{gray!20} \textbf{HLPS soft} & 1.5 &  9.845  & 4.37 & \cmark\\
			\hline
			\cellcolor{gray!20} \textbf{HLPS int.} &  1.5 & 9.854  & 4.06& \cmark \\
			\hline
			\cellcolor{gray!20} \textbf{HLPS stiff} & 1.5 & 9.93 &  3.55 &\xmark\\
			\hline
			\cellcolor{gray!20} \textbf{SLy4} & 1.5 &  9.936 & 4.01 & \cmark \\
			\hline
			\cellcolor{gray!20} \textbf{HLPS soft} & 1.8 & 9.89  &  5.185 & \xmark\\
			\hline
			\cellcolor{gray!20} \textbf{HLPS int.} & 1.8 & 9.921 & 4.381 & \cmark \\
			\hline
			\cellcolor{gray!20} \textbf{HLPS stiff} &  1.8 &  9.696 & 3.269 & \xmark\\
			\hline
			\cellcolor{gray!20}\textbf{SLy4} &  1.8 &  10.02 &  4.471 & \cmark \\
			\hline
		\end{tabular}

        \begin{tabular}{|c|c|c|c|c|}
			\hline
			\rowcolor{gray!20} \cellcolor{gray!70}\textcolor{white}{\textbf{V-QCD 5b}} & $n_t$ & $b_1$ & $c_b$ & Data\\
			\hline
			\cellcolor{gray!20} \textbf{HLPS soft} &  1.2 & 12.435 &  3.43 & \xmark\\
			\hline
			\cellcolor{gray!20} \textbf{HLPS int.} &  1.2 & 12.43 &  3.39 &\xmark\\
			\hline
			\cellcolor{gray!20} \textbf{HLPS stiff} & 1.2 &  12.51  &  3.18&\xmark \\
			\hline
			\cellcolor{gray!20} \textbf{SLy4} & 1.2 & 12.50 &  3.29 & \xmark\\
			\hline
			\cellcolor{gray!20} \textbf{HLPS soft} & 1.5 &  12.484  & 4.24 & \cmark\\
			\hline
			\cellcolor{gray!20} \textbf{HLPS int.} &  1.5 & 12.495  & 3.93& \cmark\\
			\hline
			\cellcolor{gray!20} \textbf{HLPS stiff} & 1.5 & 12.578&  3.43 &\xmark\\
			\hline
			\cellcolor{gray!20} \textbf{SLy4} & 1.5 &  12.59 & 3.876&\cmark\\
			\hline
			\cellcolor{gray!20} \textbf{HLPS soft} & 1.8 & 12.537 &  5.023 & \xmark\\
			\hline
			\cellcolor{gray!20} \textbf{HLPS int.} & 1.8 & 12.57 & 4.23 & \cmark \\
			\hline
			\cellcolor{gray!20} \textbf{HLPS stiff} &  1.8 &  12.297 &3.13 & \xmark\\
			\hline
			\cellcolor{gray!20}\textbf{SLy4} &  1.8 &  12.68 &  4.32 & \cmark \\
			\hline
		\end{tabular}}
		\caption{Summary of the EOSs and parameters for matching procedure (a).}
		\label{tab:parameters}
	\end{table}

The introduction of isospin asymmetry is a direct improvement to the level of detail described within the already successful V-QCD model. There are, however, many other refinements that could be performed, but we would like to address two:
\begin{itemize}
\item Introduction of new phases: In this work we adopted only one configuration for baryonic matter (and a similar one for quark matter), corresponding to the simplest case of a homogeneous distribution. It is not clear at what density this configuration becomes reliable, and it is expected that at lower density this approximation breaks down in favor of a spatially modulated configuration (possibly a lattice of baryons). While the inclusion of such phases may not be extremely relevant for procedure (a) since we transition to the holographic model only at higher densities, it may instead be crucial to obtain more realistic results within procedure (b) (or any other construction that employs the holographic model at lower densities). Moreover, a finite baryon number density may trigger an additional modulated instability driven by the TCS term~\cite{Domokos:2007kt,Nakamura:2009tf,Demircik:2024aig}, which may further complicate the phase diagram.
\item Inclusion of the backreaction of baryons: In this work we considered the backreaction of quarks onto the geometry, but neglected it for the baryonic phase. While this approximation is reasonable, considering that the quark phase appears at much higher density than the baryonic one, it would be ideal to treat both phases on equal footings. A possible result of this improvement is a change in the position of the transition from baryonic to quark matter, that would have repercussions on the MR curves of neutron stars, in particular it could reduce the mass $M_{\rm TOV}$ of the heaviest stable stars.
\end{itemize}
These and other improvements are for now left for future works.

\subsection*{Acknowledgments}

The work of L.~B.~is supported by the National Natural Science Foundation of China Youth Grant (Grant no.~12405084).
S.~B.~G.~thanks the Outstanding Talent Program of Henan University for partial support.
M.~J.~has been supported by an appointment to the JRG Program at the APCTP through the Science and Technology Promotion Fund and Lottery Fund of the Korean Government and by the Korean Local Governments - Gyeongsangbuk-do Province and Pohang City - and by the National Research Foundation of Korea (NRF) funded by the Korean government (MSIT) (Grant no.~2021R1A2C1010834).

\appendix

\section{Vanishing contribution of the variation of \texorpdfstring{$r_c$}{r\_c} to the symmetry energy}\label{app:vanishing_r_c}

In calculating the symmetry energy we assumed that the variation of the critical position, $r_c$, of the discontinuity in the bulk induced by the variation around isospin symmetry does not contribute to the result. 
Though intuitive, because of the perturbative argument, it is worthwhile showing explicitly that this is indeed the case.
The symmetry energy is computed from the energy per baryon, which is in turn computed from the action: Introducing an infinitesimal isospin chemical potential, $\mu_I$, we perturb the symmetric configuration, and then we read off the coefficient of the term quadratic in $\mu_I$. 
We expand the Lagrangian density in unperturbed and quadratic terms:
\beq
\mathcal{L}= \mathcal{L}_0 + \mathcal{L}_2 \mu_I^2,
\eeq
with $\mathcal{L}_0$, $\mathcal{L}_2$ independent of $\mu_I$. If the Lagrangian density vanishes for $r>r_c$, then the equilibrium condition reads:
\beq
\mathcal{L}_0(r_c) + \mathcal{L}_2(r_c) \mu_I^2=0.
\eeq
We can then expand also the critical radius $r_c$ as:
\beq
r_c = r_{c0}+\delta r_c,
\eeq
with $r_{c0}$ being the critical position of the discontinuity for symmetric matter, and $\delta r_c$ being its $\mu_I$-dependent infinitesimal variation.
Imposing the equilibrium condition at the unperturbed order allows us to determine $\delta r_c$, which is also quadratic in $\mu_I$:
\beq
\delta r_c = -\mathcal{L}_2(r_{c0})\mu_I^2\left(\mathcal{L}'_0(r_{c0})\right)^{-1}.
\eeq
We now need to prove that the Lagrangian does not depend on $\delta r_c$ at the quadratic order in $\mu_I$. The Lagrangian is given by:
\begin{align}
L&= \int^{r_c}_{0}\d r\left(\mathcal{L}_0 + \mathcal{L}_2\mu_I^2\right)\non
&= \int^{r_{c0}}_{0}\d r\left( \mathcal{L}_0 +\mathcal{L}_2\mu_I^2\right) + \int_{r_{c0}}^{r_{c0}+\delta r_c} \d r \left( \mathcal{L}_0 +\mathcal{L}_2\mu_I^2\right)\non
&=\int^{r_{c0}}_{0}\d r\left( \mathcal{L}_0 +\mathcal{L}_2\mu_I^2\right) +\delta r_c \mathcal{L}_0 (r_{c0}) + \delta r_c \mathcal{L}_2(r_{c0}) \mu_I^2.
\end{align}
The first term is the Lagrangian with the perturbation induced only at the field level, the one we use for the computation of the symmetry energy. The second term vanishes because $\mathcal{L}_0(r_{c0})=0$ by definition of $r_{c0}$. Finally, the third term only contributes at order $\mu_I^4$.\hfill$\square$

\bibliographystyle{JHEP}
\bibliography{refs}

\end{document}